# Stochastic Constitutive Model of Isotropic Thin Fiber Networks Based on Stochastic Volume Elements

**Rami Mansour [1], Artem Kulachenko [1,*], Wei Chen [2] and Mårten Olsson [1]**

[1] Department of Solid Mechanics, Royal Institute of Technology, SE-100 44 Stockholm, Sweden; ramimans@kth.se (R.M.); mart@kth.se (M.O.)
[2] Department of Mechanical Engineering, Northwestern University, Evanston, IL 60208, USA; weichen@northwestern.edu
* Correspondence: artem@kth.se; Tel.: +46-8-790-89-44

**Abstract:** Thin fiber networks are widely represented in nature and can be found in man-made materials such as paper and packaging. The strength of such materials is an intricate subject due to inherited randomness and size-dependencies. Direct fiber-level numerical simulations can provide insights into the role of the constitutive components of such networks, their morphology, and arrangements on the strength of the products made of them. However, direct mechanical simulation of randomly generated large and thin fiber networks is characterized by overwhelming computational costs. Herein, a stochastic constitutive model for predicting the random mechanical response of isotropic thin fiber networks of arbitrary size is presented. The model is based on stochastic volume elements (SVEs) with SVE size-specific deterministic and stochastic constitutive law parameters. The randomness in the network is described by the spatial fields of the uniaxial strain and strength to failure, formulated using multivariate kernel functions and approximate univariate probability density functions. The proposed stochastic continuum approach shows good agreement when compared to direct numerical simulation with respect to mechanical response. Furthermore, strain localization patterns matched the one observed in direct simulations, which suggests an accurate prediction of the failure location. This work demonstrates that the proposed stochastic constitutive model can be used to predict the response of random isotropic fiber networks of arbitrary size.

## 1. Introduction

Materials are characterized by certain degrees of random variations in their mechanical properties. This is true for all materials but especially pronounced in disordered materials such as thin fiber networks [1]. Such variations can be the cause of unexplained occasional failures that cannot be predicted by deterministic material models [2–4]. It is, therefore, crucial to develop a sound stochastic approach in studying mechanical failure of thin fiber networks of arbitrary size.

The rapid development of characterization tools enables the quantification of randomness at different scales and the construction of random realizations of a fiber network. In our previous works, the mechanical behavior of randomly generated networks was investigated using detailed direct micromechanical simulations [5–8] (see Section 2.1). However, although such simulations can capture the complicated mechanisms of failure, they cannot yet be employed for product development due to the overwhelming computational costs required to capture the relevant product sizes. For instance, simulation of the uniaxial mechanical response up to strain localization and failure of a 24 mm $\times$ 24 mm fiber network takes two days on a modern 28-core, 128 GB RAM supercomputer.

Therefore, the straightforward use of direct micromechanical simulations for predicting the mechanical response of large structures of complex geometry will be impossible in the foreseeable future. In Figure 1, examples of paper-based products composed of random fiber networks are shown. A micro-tomography of a typical fiber network (Figure 1b) followed by numerical reconstruction (Figure 1c) is also shown. An example of random occurrence of breaks of paper-based products can be found in paper-making machines. Every break costs around 6000 € and some paper machines can experience up to six breaks a day. With the speed of the machine reaching 2000 m/min, the break can, therefore, occur once per 480 km of produced paper, which makes it an extremely stochastic event.

The aforementioned difficulties associated with predicting the mechanical response of larger structures based on random local material properties has been the topic of extensive research for a wide variety of materials including fiber systems [9–17]. In recent works, different methods based on stochastic volume elements (SVEs) [18] have been used in a variety of applications to alleviate the computational burden. In Reference [19], microstructure–property relations of heterogeneous materials were assessed using a hierarchical decomposition of statistically representative volume elements (RVEs) [20] into smaller SVEs. In Reference [21], analysis of particle reinforced viscoelastic polymer nanocomposites was also performed using SVEs. Similarly, a stochastic multiscale model for polycrystalline materials based on SVEs was developed in Reference [22]. In general, three different scales are involved in such analysis: (1) the micro-scale, which is a characteristic size of the microstructure; (2) the mesoscale, an intermediate scale such as the size of the volume element over which a homogenization can be performed; and (3) the macro-scale which is the size of the structural problem.

In this work, a stochastic constitutive model is proposed to address the aforementioned computational difficulties associated with modeling large random fiber networks. The model is based on SVEs and random field generation of local material properties. It includes three major components: a deterministic constitutive model, characterization of SVE size-dependent model parameters, and generation of two correlated non-Gaussian random spatial fields describing local material properties. The model is numerically validated by ensuring (1) accurate prediction of mechanical failure, i.e., strength and strain to failure as well as strain localization pattern; and (2) generation of continuum random realizations that are statistically equivalent to the ones generated using a direct simulation of fiber networks.

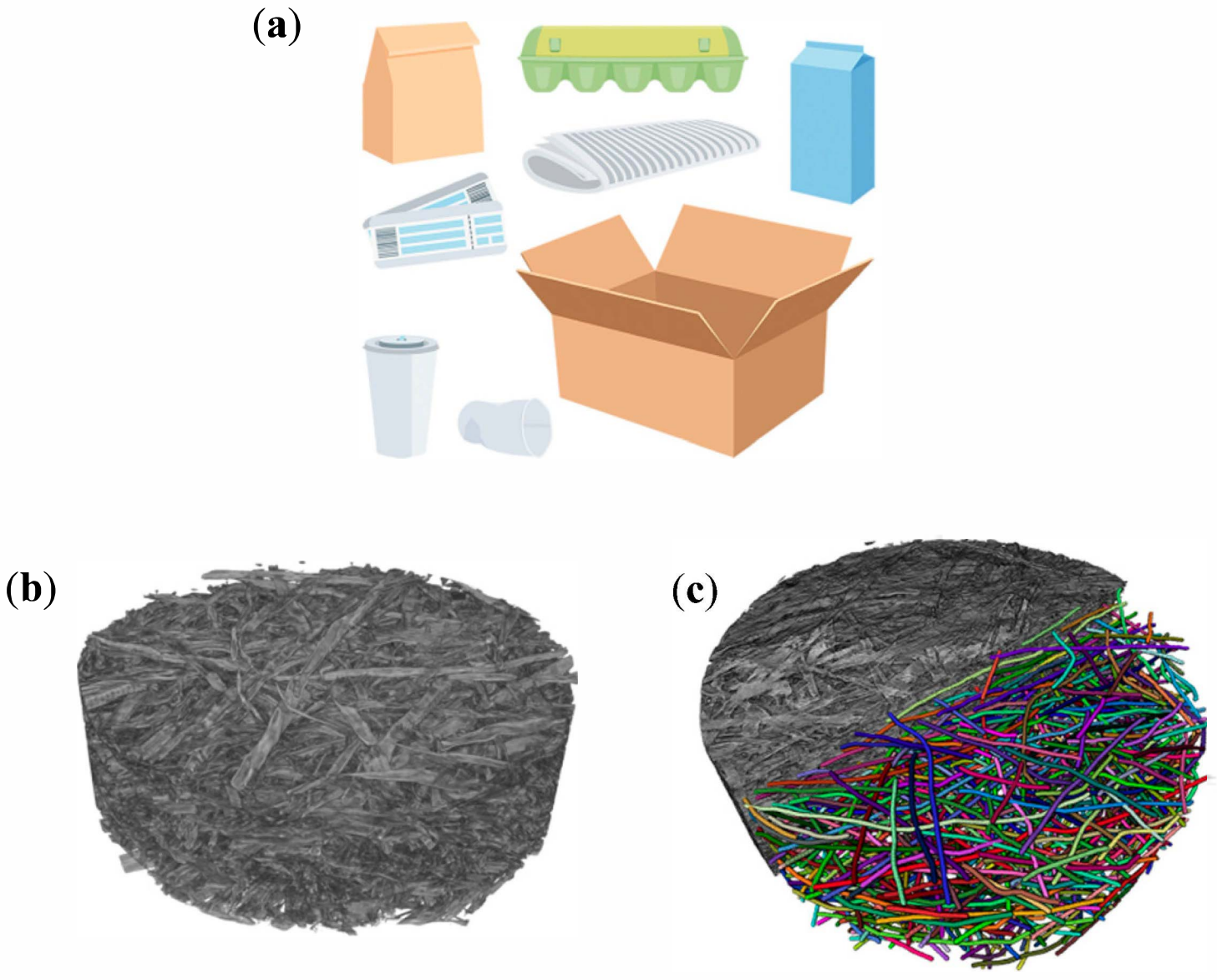

**Figure 1.** (**a**) Example of fiber-based product, (**b**) micro-tomography of fiber network, and (**c**) fiber network schematic reconstruction.

The methodology proposed in this work is generic but demonstrated for isotropic fiber networks subjected to uniaxial loading until failure. The paper is organized as follows. An introduction to the direct fiber network simulation as well as important concepts in stochastic multiscale modeling and random spatial field modeling are presented in Section 2. The proposed stochastic constitutive model is presented in Section 3. In Section 4, a validation of the model is first performed for small specimens by comparing the results to direct fiber network simulations. The approach is thereafter demonstrated by random generation and mechanical simulation of large continuum models of fiber networks.

## 2. Technical Background

### *2.1. Direct Simulation of a Thin Fiber Network*

A 3D network of fibers is generated using a deposition technique in which the fibers are sequentially deposited on a flat surface from two sides. The deposition algorithm can be outlined as follows:

1. The fiber geometry is chosen from the fiber characterization data acquired with FiberLab (Valmet Fiber Image Analyzer), which is an apparatus for measuring fiber characteristics. It contains length, width, height, wall thickness, and curvature. The curvature is represented through an arc of constant curvature located in a single plane parallel to the deposition plane. The cross-sectional data is corrected using microtomography scans of a paper produced using the considered softwood kraft pulp. The details of the correction are described elsewhere [6].
2. The fiber orientation is chosen randomly in this work although it can be controlled to match a specific distribution.
3. The fiber position before deposition onto the domain is chosen randomly.
4. The first fibers are deposited on the flat plane consecutively from above or below. For the subsequent fibers, we first find the intersection between them and the previously deposited fibers in the plane (Figure 2a).
5. The found intersection points are lifted discretely to exclude penetration (Figure 2b). The contact search diameter depicted in the figure corresponds to the height of the fiber, which is smaller or equal to the width of the fiber normally.
6. The fiber geometry is smoothed to remove discontinuities caused by the previous step (Figure 2c). During the smoothing, we control the maximum angle the fibers can form, and it was set to 5 degrees in this study.
7. When the grammage (the basis weight or the weight per unit area) of the network has reached the prescribed value, the deposition procedure is stopped. The grammage used in this work was 28 $g/m^2$, which is relatively low but corresponds to the set of handsheets used to calibrate the measurements in Reference [6].
8. The thickness of the network is evaluated and measured using the procedure described in Reference [6]. Thereafter, the thickness is brought to the target value by uniform scaling of the coordinates in the thickness direction with respect to the center plane of the network. The target value in this study was 68 micrometers, which corresponds to the measured value used in the calibration [6]. The scaling may result in interpenetration, which are zeroed out during the subsequent computations.

After generation of the geometry, the fibers are represented with a fine mesh of curved segments and imported into a custom finite element code. The fibers are converted into cubic splines and a full 3D network finite element model [5,6,23] is used, in which fibers are resolved as a chain of 3D quadratic Timoshenko/Reissner beam elements [24] with six degrees of freedom at each node. The implicit time-integration is used. During contact detection, cross-sections of the fibers are treated as circular and rigid, with the diameter equal to the mean of the values of fiber width and height. The mechanical bonds behavior is described with traction-separation laws with a cohesive zone model based on the contact forces [6].

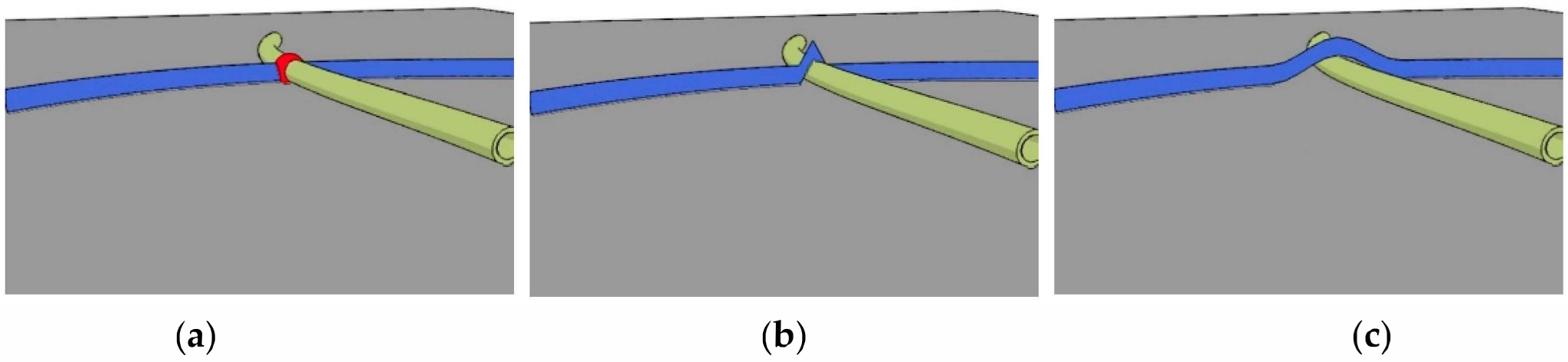

**Figure 2.** Steps in deposition procedure: (**a**) finding intersections with previously deposited fibers, (**b**) lifting the intersection point, and (**c**) smoothing the fiber.

Table 1 summarizes the details of the fibers used in the analysis. The underlying data is identical to that used by Reference [6], who focused on the mean values of the tensile properties of the networks. The constitutive response of the fibers is described by a bilinear plasticity model with the material parameters as listed in Table 2.

**Table 1.** Fiber geometrical data used in the network simulation, based on the direct measurement on wet pulp (FMA), on dry sheets (μCT), and numerical parameters in terms of length-weighted mean and standard deviation (SD) values.

| | **Mean** | **SD** |
|---|---|---|
| Fiber length, mm | 2.34 | 0.90 |
| Fiber width, μm | 23.83 | 7.09 |
| Fiber wall thickness, μm | 3.96 | 1.90 |
| WH ratio, (-) | 2.9 | 1.72 |
| Fiber shape factor | 0.945 | 0.015 |
| Maximum interface angle, ° | 5 | - |
| Radius swelling factor [1], (-) | 0.78 | 0.68 |
| Wall thickness swelling factor, (-) | 0.528 | 0.31 |

[1] The ratio between dry and wet measured radius and wall thickness, respectively.

**Table 2.** Fiber material parameters used in the network simulation.

| **Elastic Modulus, GPa** | **Tangent Modulus, GPa** | **Yield Stress, MPa** |
|---|---|---|
| 30 | 10 | 150 |

The contact conditions at the fiber bonds are governed by a bilinear cohesive traction–separation law. It requires the definitions of a bond's stiffness, a bond's strength and separation, see Table 3. Since the beam's cross-section is rigid against the local normal forces and shear forces, the physical compliance of the fiber at the bonding sites is represented solely with the stiffness of the penalty-based contact element.

**Table 3.** Characteristics of bonds used in the network simulation.

| | **Tangential Direction** | **Normal Direction** |
|---|---|---|
| Bond strength, mN | 11.00 | 2.75 |
| Bond stiffness, $10^9$ N/m | 8.90 | 8.00 |
| Separation distance, μm | 1.56 | 0.35 |

The selection of the parameters listed in Tables 2 and 3 are based on the experimental calibration in the tensile test described in Reference [6].

The software used for the fiber network reconstruction as well as the input data are uploaded as a Supplementary Materials to the current paper.

### 2.2. Multi-Scale Modeling

Multi-scale mechanical modeling of thin fiber networks is a powerful technique aimed at predicting the mechanical response at the macro-mechanical level, based on properties at the meso- and microscale. The up-scaling methodology in multi-scale methods generally relies on either SVEs [18] or RVEs [25–29]. In the former, elements are characterized by a stochastic variation in their mechanical response. Put another way, SVEs taken from different locations within the macroscopic body have different material properties due to random variation at the micromechanical level. This is opposed to the concept of RVEs, for which the material properties or any chosen statistical descriptor remain constant regardless of the location of the RVEs within the macroscopic body [20]. An RVE can, therefore, be defined as the smallest volume over which statistical representation can be made for the considered material property [29]. It is used to determine the corresponding effective properties for a homogenized macroscopic model. According to the micro–meso–macro principle [30], the RVE should be sufficiently large in order to contain representative information about the microstructure, but its size should be small compared to the studied macroscopic body.

Multi-scale strategies aiming at reconstructing a macroscopic body using RVEs may not be successful, since an RVE may not exist [29]. An attempt to find an appropriate size of the RVE can be made by increasing the size of an SVE until its response becomes independent of its spatial location. In statistical terms, an appropriate RVE size is considered to be found when the standard deviation of the chosen material property vanishes, and its mean value converges to a constant value. It has been shown that, for instance, when considering a statistical descriptor related to the softening behavior of the material in question, its mean value does not converge with increasing element size [5]. In such cases, a multi-scale model based on the response of SVEs may be more appropriate.

As opposed to multi-scale applications where the focus is solely on the linear elastic response, prediction of strength to failure of thin fiber networks is the main target of the present work. In Figure 3, a multi-scale model for thin fiber networks based on SVEs is described. A fiber network is spatially discretized using a chosen SVE size, see Figure 3a. It should be noted that the discretization may involve overlapping SVEs. In Figure 3b, the response from direct numerical simulations on SVEs from different spatial locations is shown and the random variation is apparent. In order to correctly model the transition between different scales, the fiber network is created using a statistically equivalent microstructure, Figure 3c. However, direct numerical simulation on a large fiber network in Figure 3a is not possible due to formidable computational cost. The aim of this multi-scale strategy is therefore to predict the mechanical response of the fiber network based on the mechanical response of the SVEs in Figure 3b.

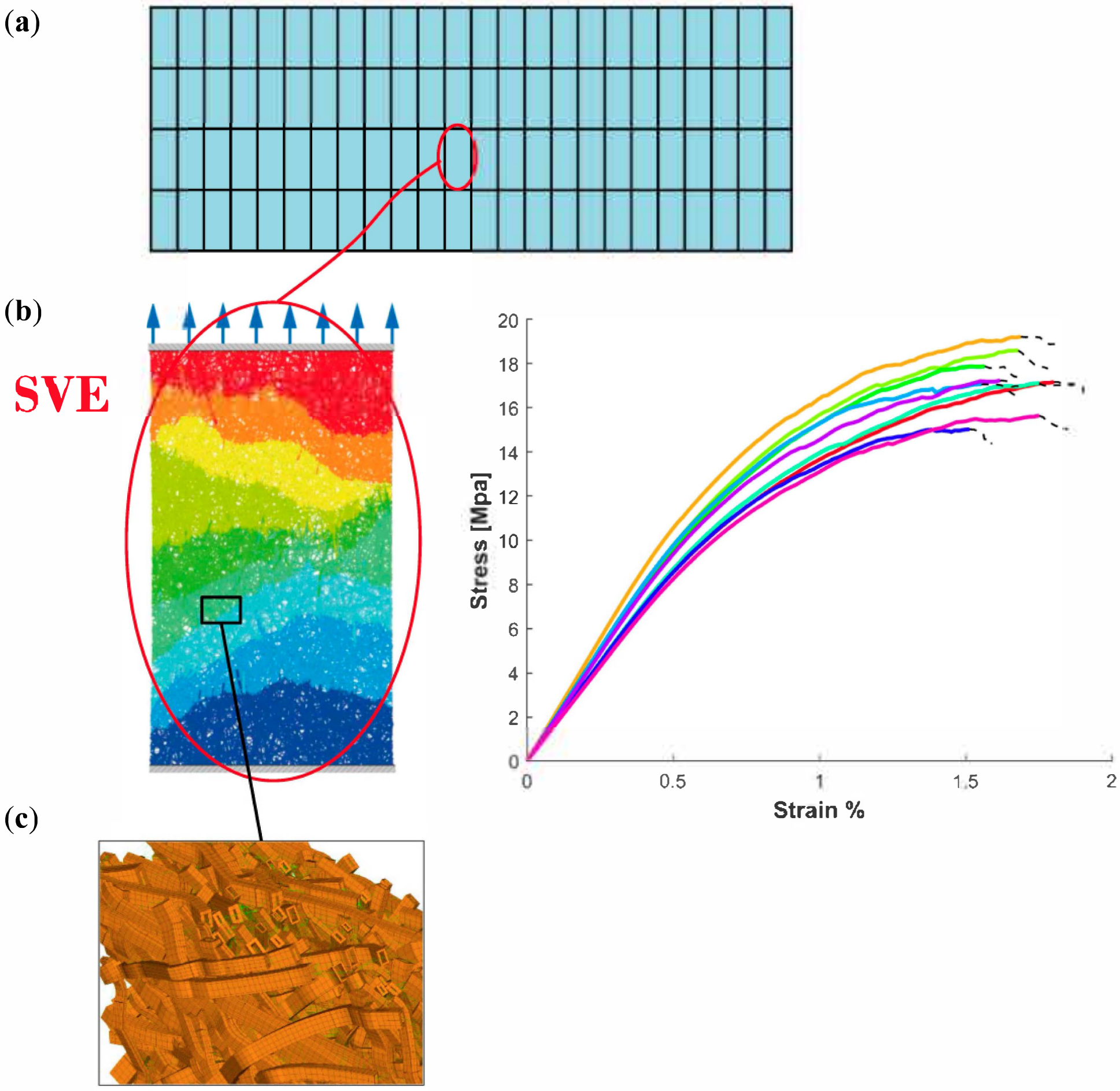

**Figure 3.** (**a**) A large paperboard. (**b**) Numerical testing of a uniaxially loaded Stochastic Volume Element (SVE) [7]. (**c**) Computer simulation of a statistically equivalent fiber network.

### *2.3. Random Field Representation*

Accurate models of random local spatial variations of material response, such as the one observed in Figure 3b, are necessary to correctly predict mechanical failure of the disordered fiber network. In order to describe these local variations, random spatial fields are needed. A random field can be represented as an infinite expansion of orthogonal basis functions and random expansion coefficients. Let $h(\boldsymbol{r}, \omega) : \mathcal{D} \times \Omega \rightarrow \mathbb{R}$ be such a random continuous spatial field defined over a spatial domain $\mathcal{D}$, where $\boldsymbol{r} = \begin{bmatrix} x & y & z \end{bmatrix}^T$ is the spatial vector and $\omega \in \Omega$ denotes an element of the sample space indicating that the involved quantity is random. The random field's covariance function is denoted as $C(\boldsymbol{r}, \boldsymbol{r}') = \operatorname{cov}(h(\boldsymbol{r}, \omega), h(\boldsymbol{r}', \omega))$. By the definition of a covariance function, it has the spectral decomposition [31]

$$C(\boldsymbol{r}, \boldsymbol{r}') = \sum_{\boldsymbol{i}=\mathbf{1}}^{\infty} \lambda_i \eta_i(\boldsymbol{r}) \eta_i(\boldsymbol{r}'), \tag{1}$$

where $\lambda_i$ and $\eta_i(\boldsymbol{r})$ are the eigenvalues and eigenfunctions of the covariance function, respectively. That is, they are the solution to the Fredholm integral

$$\int_{\mathcal{D}} C(\boldsymbol{r}, \boldsymbol{r}') \eta_i(\boldsymbol{r}) d\boldsymbol{r} = \lambda_i \eta_i(\boldsymbol{r}'). \tag{2}$$

The Karhunen-Loeve (KL) expansion [31] of the random field $h(\boldsymbol{r}, \omega)$ uses the eigenfunctions of the covariance function as expansion basis according to

$$h(\boldsymbol{r},\omega)=\bar{h}(\boldsymbol{r})+\sum_{i=1}^{\infty}\sqrt{\lambda_i}\eta_i(\boldsymbol{r})\xi_i(\omega), \tag{3}$$

where $\bar{h}(\boldsymbol{r})$ is the mean function and $\xi_i(\omega)$ are orthogonal random variables with zero mean and unit variance. Explicit expression for $\xi_i(\omega)$ can be found by multiplying Equation (3) by $\eta_i(\boldsymbol{r})$ and integrating over the spatial domain $\mathcal{D}$ according to

$$\xi_i(\omega)=\frac{1}{\sqrt{\lambda_i}}\int_{\mathcal{D}}\Big[h(\boldsymbol{r},\omega)-\bar{h}(\boldsymbol{r})\Big]\eta_i(\boldsymbol{r})d\boldsymbol{r}, \tag{4}$$

where use is made of the orthogonality property of the set $\{\eta_i(\boldsymbol{r})\}_{i=1}^{\infty}$.

The choice of basis function and expansion coefficients in the KL expansion is motivated by truncating the infinite expansion according to Equation (3) and minimizing the total truncation error. If the expansion is truncated at the $M$-th term, the truncation error is given by $e_M(\boldsymbol{r}) = \sum_{i=M+1}^{\infty}\sqrt{\lambda_i}\eta_i(\boldsymbol{r})\xi_i(\omega)$. It can be shown that the total mean square error $\int_{\mathcal{D}} e_M(\boldsymbol{r})^2 d\boldsymbol{r}$ is minimized (subject to the orthogonality condition of $\{\eta_i(\boldsymbol{r})\}$) if the Fredholm integral equation according to Equation (2) is satisfied. Therefore, of all possible choices of expansion basis, the Karhunen–Loeve (KL) expansion [31] is the best approximation of the original random field in the sense that it minimizes the total mean-square error resulting of its truncation.

The usefulness of the Karhunen–Loeve expansion hinges on the ability to determine the eigenvalues and eigenfunctions of the random field covariance function through the Fredholm integral equation. In this work, the continuous random field is discretized at $N$ spatial points $\boldsymbol{r}_1, \boldsymbol{r}_2, \ldots, \boldsymbol{r}_N$ in the domain $\mathcal{D}$. An $N \times N$ covariance matrix with elements $\boldsymbol{C}_{pq} = C(\boldsymbol{r}_p, \boldsymbol{r}_q)$ can be defined, where $\boldsymbol{r}_p$ and $\boldsymbol{r}_q$ are two spatial locations. For this discrete and finite case, the Fredholm integral equation can be rewritten as

$$\boldsymbol{C}\boldsymbol{\eta}_i=\lambda_i\boldsymbol{\eta}_i, \tag{5}$$

where $\{\lambda_i\}_{i=1}^{N}$ and $\{\boldsymbol{\eta}_i\}_{i=1}^{N}$ are eigenvalues and eigenvectors of the covariance matrix $\boldsymbol{C}$. That is, when the random field is discretized, operations on functions are transformed into operation on matrices. The KL expansion of the discrete random field [32–35] can be written as

$$\boldsymbol{h}(\omega)=\bar{\boldsymbol{h}}+\sum_{i=1}^{N}\sqrt{\lambda_i}\boldsymbol{\eta}_i\xi_i(\omega), \tag{6}$$

where

$$\boldsymbol{h}(\omega)=\begin{bmatrix} h(\boldsymbol{r}_1,\omega) & h(\boldsymbol{r}_2,\omega) & \cdots & h(\boldsymbol{r}_N,\omega)\end{bmatrix}^T \tag{7}$$

denotes a random vector whose elements are the random field values at the $N$ discrete points and $\bar{\boldsymbol{h}}=\begin{bmatrix} \bar{h}(\boldsymbol{r}_1) & \bar{h}(\boldsymbol{r}_2) & \cdots & \bar{h}(\boldsymbol{r}_N)\end{bmatrix}^T$. The random coefficients are obtained from

$$\xi_i(\omega)=\frac{1}{\sqrt{\lambda_i}}\left(\boldsymbol{h}(\omega)-\bar{\boldsymbol{h}}\right)^T\boldsymbol{\eta}_i. \tag{8}$$

In general, the underlying covariance function $C(\boldsymbol{r},\boldsymbol{r}')$ is not known. One crucial task is therefore the modelling of the covariance function based on sampling and characterization of the spatial random fields. It should be noted that the above description can be extended to represent multivariate random spatial fields. In this case, the covariance matrix $\boldsymbol{C}$ is composed of both auto- and cross- covariances, where the latter describes the correlation between different spatial fields, see Section 3.2.

## 3. Continuum Random Realization and Mechanical Failure of Thin Fiber Networks

### *3.1. Overview of Methodology*

The aim of this work is to model computationally expensive random isotropic 3D-fiber network realizations by statistically equivalent random continuum realizations, see Figure 4. The model is based on the random spatial fields of strain to failure $\varepsilon_f(\boldsymbol{r}, \omega)$ and strength $\sigma_f(\boldsymbol{r}, \omega)$ which are found using direct mechanical simulation on SVEs. The latter is defined as the ratio of the applied force on the SVE and the average thickness of the whole fiber network. The characterization and reconstruction of the random spatial fields as well as the stochastic constitutive model used to predict the mechanical response of isotropic fiber networks of arbitrary size, is presented in Sections 3.2–3.5.

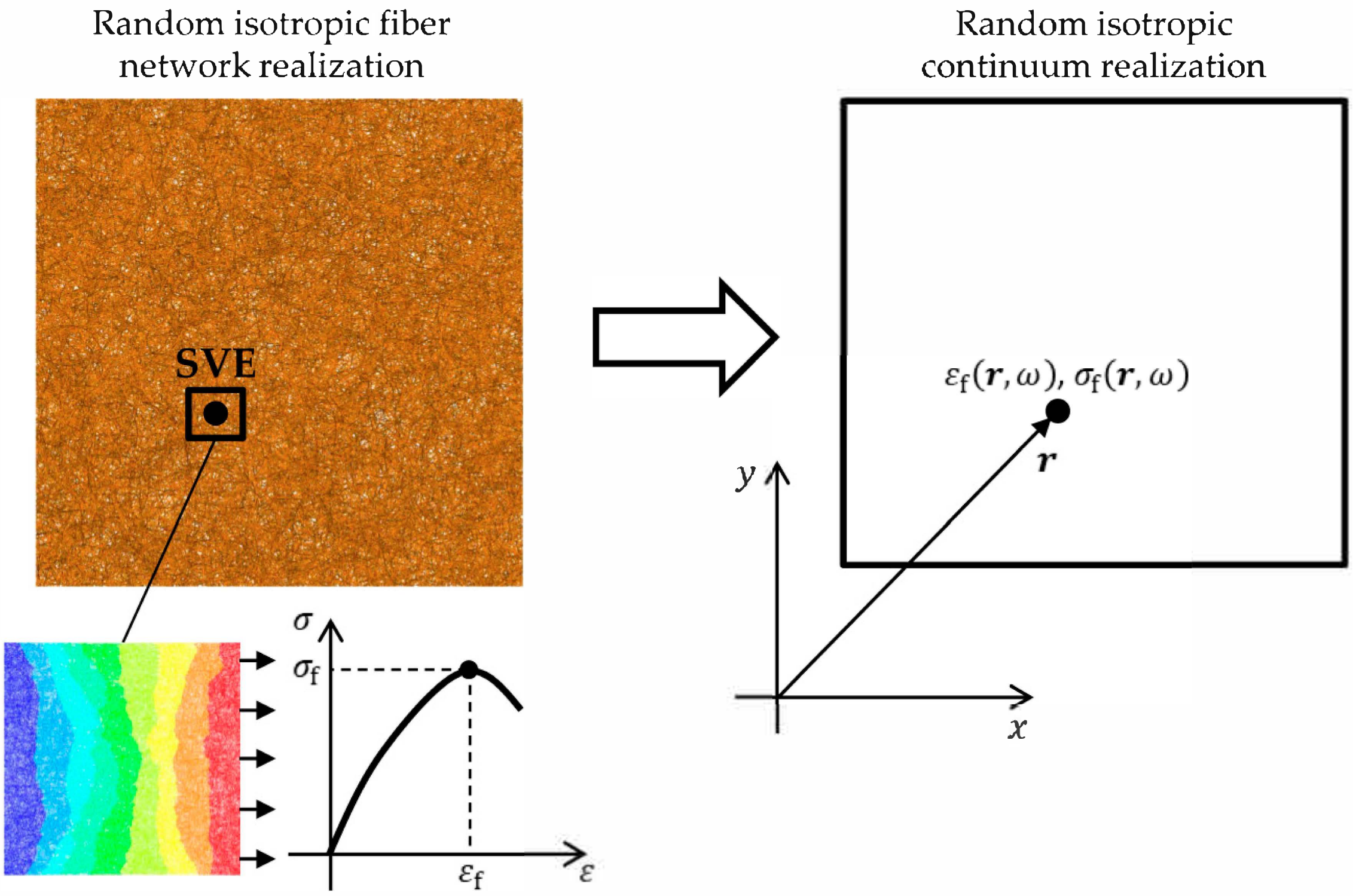

**Figure 4.** A random realization of a fiber network using a deposition technique and an equivalent continuum realization based on random spatial fields of strain and strength to failure.

### *3.2. Characterization and Simulation of Random Spatial Fields of Strength and Strain to Failure Based on Stochastic Volume Elements (SVEs)*

The proposed continuum model of isotropic 3D fiber networks is based on two continuous macro-mechanical spatial fields describing the material properties at each material point in a 2D spatial domain $\boldsymbol{r} = \begin{bmatrix} x & y \end{bmatrix}^T$. These are the uniaxial strength $\sigma_f(\boldsymbol{r}, \omega)$ and uniaxial strain to failure $\varepsilon_f(\boldsymbol{r}, \omega)$. The characterization methodology of the random spatial fields is detailed in Table 4.

The first step is the sampling of the spatial fields. Since the fiber network is isotropic, it is sufficient to determine the spatial fields in one coordinate direction, see Figure 5a. A specimen of length $L$ is discretized using SVEs of size $L_{SVE} \times W_{SVE}$. The discretization distance $\Delta L$ is the distance between the centers of two neighboring SVEs corresponding to material points in the continuum model. All SVEs are cut from the fiber network and a numerical uniaxial test is performed. The resulting one-dimensional spatial fields of the strain to failure $\varepsilon_f(x, \omega)$ and strength $\sigma_f(x, \omega)$ are schematically shown in Figure 5b.

An important characteristic of these spatial fields is their mutual correlation. It is computed from the Pearson's correlation coefficient based on the one-dimensional spatial fields in Figure 5b according to

$$\rho^{t}_{\varepsilon_{\mathrm{f}}\sigma_{\mathrm{f}}} = \frac{N\sum_{i=1}^{N}\varepsilon_{\mathrm{f}}(x_i)\sigma_{\mathrm{f}}(x_i) - \sum_{i=1}^{N}\varepsilon_{\mathrm{f}}(x_i)\sum_{i=1}^{N}\sigma_{\mathrm{f}}(x_i)}{\sqrt{N\sum_{i=1}^{N}\varepsilon_{\mathrm{f}}(x_i)^2 - \left(\sum_{i=1}^{N}\varepsilon_{\mathrm{f}}(x_i)\right)^2}\sqrt{N\sum_{i=1}^{N}\sigma_{\mathrm{f}}(x_i)^2 - \left(\sum_{i=1}^{N}\sigma_{\mathrm{f}}(x_i)\right)^2}}. \tag{9}$$

In Equation (9), superscript "t" denotes that this is a target value that needs to be satisfied when simulating new realization of $\varepsilon_{\mathrm{f}}(x,\omega)$ and $\sigma_{\mathrm{f}}(x,\omega)$ or $\varepsilon_{\mathrm{f}}(\boldsymbol{r},\omega)$ and $\sigma_{\mathrm{f}}(\boldsymbol{r},\omega)$.

**Table 4.** Characterization and simulation methodology.

| | Steps | Details |
|---|---|---|
| **Characterization** | 1D Sampling of: $\varepsilon_{\mathrm{f}}(x,\omega)$ and $\sigma_{\mathrm{f}}(x,\ \omega)$ | Figure 5a,b |
| | Computing Pearson's correlation coefficients: $\rho^{t}_{\varepsilon_{\mathrm{f}}\sigma_{\mathrm{f}}}$ | Equation (9) |
| | Fitting probability distribution of: $\varepsilon_{\mathrm{f}}(\omega)$ and $\sigma_{\mathrm{f}}(\omega)$ | Equation (10)<br>Figure 5c |
| | Transformation to correlated Gaussian fields: $g_1(x,\omega)$ and $g_2(x,\ \omega)$ | Equation (11)<br>Figure 5d |
| | Average distance between two zero level upcrossings: $\mu^{t}_{\mathrm{upcr1}}$ and $\mu^{t}_{\mathrm{upcr2}}$ | Equations (12) and (13)<br>Figure 11 b,c |
| | Modeling of auto- and cross-covariance function for multivariate Gaussian fields: $K_{11}(\boldsymbol{r},\boldsymbol{r}'\vert\ell_1)$, $K_{22}(\boldsymbol{r},\boldsymbol{r}'\vert\ell_2)$, $K_{12}(\boldsymbol{r},\boldsymbol{r}'\vert\rho_{12},\ell_{12})$ | Equation (14)<br>Equation (15) |
| | Determination of constants: $\ell_1,\ell_2,\rho_{12},\ell_{12}$<br>Initial guesses: $\ell_1^{(0)},\ell_2^{(0)},\rho_{12}^{(0)},\ell_{12}^{(0)}$ | Equation (19)<br>Equation (20) |
| **Simulation** | Simulation using Karhunen-Loeve expansion of 2D correlated Gaussian fields: $g_1(\boldsymbol{r},\omega)$ and $g_2(\boldsymbol{r},\ \omega)$ | Equation (21) |
| | Transformation to original random field: $\varepsilon_{\mathrm{f}}(\boldsymbol{r},\omega)$ and $\sigma_{\mathrm{f}}(\boldsymbol{r},\omega)$ | Equation (22) |

The next step is to fit a probability density function for the random variables $\varepsilon_{\mathrm{f}}(\omega)$ and $\sigma_{\mathrm{f}}(\omega)$. In Figure 5c, histograms of $m$ random outcomes of $\varepsilon_{\mathrm{f}i}$ and $\sigma_{\mathrm{f}i}$ are shown. These are generated by cutting SVEs from random spatial locations in a large fiber network. The probability density functions (pdf) $f_{\varepsilon_{\mathrm{f}}}(\varepsilon_{\mathrm{f}})$ and $f_{\sigma_{\mathrm{f}}}(\sigma_{\mathrm{f}})$ are approximated based on the kernel density estimators

$$\begin{cases} f_{\varepsilon_{\mathrm{f}}}(\varepsilon_{\mathrm{f}}) = \frac{1}{mh_{\varepsilon_{\mathrm{f}}}}\sum_{i=1}^{m}\varphi\left(\frac{\varepsilon_{\mathrm{f}i}-\varepsilon_{\mathrm{f}}}{h_{\varepsilon_{\mathrm{f}}}}\right) \\ f_{\sigma_{\mathrm{f}}}(\sigma_{\mathrm{f}}) = \frac{1}{mh_{\sigma_{\mathrm{f}}}}\sum_{i=1}^{m}\varphi\left(\frac{\sigma_{\mathrm{f}i}-\sigma_{\mathrm{f}}}{h_{\sigma_{\mathrm{f}}}}\right) \end{cases}, \tag{10}$$

where $\varphi$ is the standard normal density function and $h_{\varepsilon_{\mathrm{f}}}, h_{\sigma_{\mathrm{f}}} > 0$ are smoothing bandwidth.

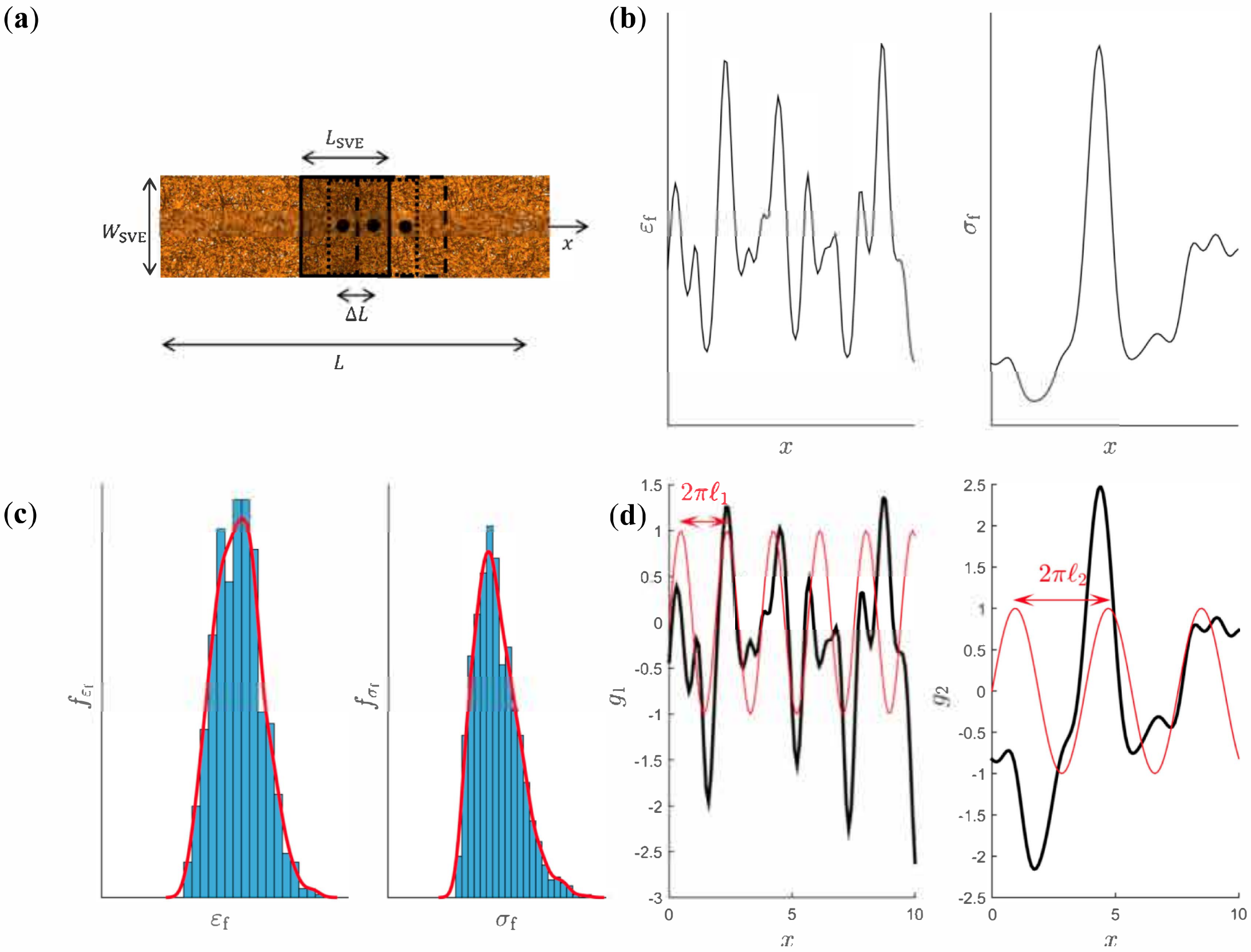

**Figure 5.** (**a**) Sample used to characterize the spatial random fields of $\varepsilon_{\mathrm{f}}$ and $\sigma_{\mathrm{f}}$ with discretization distance $\Delta L$. Schematic (**b**) Spatial random fields of $\varepsilon_{\mathrm{f}}$ and $\sigma_{\mathrm{f}}$. (**c**) Univariate approximation of the probability density function of $\varepsilon_{\mathrm{f}}$ and $\sigma_{\mathrm{f}}$. (**d**) Transformed Gaussian random fields.

Thereafter, a transformation from the non-Gaussian fields ($\varepsilon_{\mathrm{f}}(x,\omega)$ and $\sigma_{\mathrm{f}}(x,\omega)$) to correlated Gaussian random fields ($g_1(x,\omega)$ and $g_2(x,\omega)$) with zero mean and unit variance is performed. It should be emphasized that this step is performed to facilitate the modeling of the covariance function and that the Gaussian fields are later transformed back to the original spatial fields. The transformation to Gaussian random fields is performed using the relation [36,37]

$$\begin{cases} g_1(x,\omega) = \phi^{-1}\{F_{\varepsilon_{\mathrm{f}}}[\varepsilon_{\mathrm{f}}(x,\omega)]\} \\ g_2(x,\omega) = \phi^{-1}\{F_{\sigma_{\mathrm{f}}}[\sigma_{\mathrm{f}}(x,\omega)]\} \end{cases}, \tag{11}$$

where $F_{\varepsilon_{\mathrm{f}}}(\varepsilon_{\mathrm{f}})$ and $F_{\sigma_{\mathrm{f}}}(\sigma_{\mathrm{f}})$ are the cumulative density functions (CDFs) found by integration of the kernel density estimators according to Equation (10), and $\phi$ is the standard normal CDF. In Figure 5d), these Gaussian random fields are schematically shown. An important characteristic of each of these one-dimensional Gaussian fields $g_i$ is the average distance between two zero level up-crossings denoted by $\mu^t_{\mathrm{upcr}i}$ [38]. These are computed based on the number of up-crossings in the interval [0, $L$] corresponding to the specimen length in Figure 5a, defined as

$$\boldsymbol{N}_{\mathbf{upcr}\boldsymbol{i}} \triangleq \#\left\{\boldsymbol{x} \in \begin{bmatrix} \mathbf{0} & \boldsymbol{L} \end{bmatrix} : \boldsymbol{g}_{\boldsymbol{i}}(\boldsymbol{x}) = \mathbf{0}, \frac{\boldsymbol{d}\boldsymbol{g}_{\boldsymbol{i}}}{\boldsymbol{d}\boldsymbol{x}} > \mathbf{0}\right\}. \tag{12}$$

The average distance between two zero level up-crossings for $g_1$ and $g_2$ is therefore given by

$$\begin{cases} \mu^t_{\mathrm{upcr1}} = \frac{L}{N_{\mathrm{upcr1}}-1} \\ \mu^t_{\mathrm{upcr2}} = \frac{L}{N_{\mathrm{upcr2}}-1} \end{cases}. \tag{13}$$

From Equation (13) it is seen that the required specimen length $L$ has to be chosen sufficiently large so that both $\mu^t_{\mathrm{upcr1}}$ and $\mu^t_{\mathrm{upcr2}}$ converges to a constant value, see Section 4.1 and Figure 11c,d.

These values are target values in the sense that they need to be satisfied when simulating new realization of $\varepsilon_{\mathrm{f}}(x,\omega)$ and $\sigma_{\mathrm{f}}(x,\omega)$ followed by the transformation according to Equation (11).

A crucial step in the characterization of random spatial fields is to choose an appropriate model for the underlying covariance function. This is easiest performed in the transformed Gaussian random variable space. The choice of model is important in that the properties of the Gaussian random field, such as smoothness or differentiability, are determined by the eigenvalues and eigenfunctions of the covariance function. In this work, the squared exponential kernel is used to model the auto-covariance for $g_1(\boldsymbol{r},\omega)$ and $g_2(\boldsymbol{r},\omega)$, denoted by $K_{11}(\boldsymbol{r},\boldsymbol{r}')$ and $K_{22}(\boldsymbol{r},\boldsymbol{r}')$, respectively, according to

$$\begin{cases} K_{11}(\boldsymbol{r},\boldsymbol{r}') = exp\left(-\frac{|\boldsymbol{r}-\boldsymbol{r}'|^2}{2\ell_1{}^2}\right) \\ K_{22}(\boldsymbol{r},\boldsymbol{r}') = exp\left(-\frac{|\boldsymbol{r}-\boldsymbol{r}'|^2}{2\ell_2{}^2}\right) \end{cases} . \tag{14}$$

This choice of covariance function satisfies the condition of high smoothness and is infinitely differentiable. From Equation (14), it can be seen that the auto-covariance is only a function of the distance between two spatial locations $|\boldsymbol{r}-\boldsymbol{r}'|$ in relation to a characteristic length scale ($\ell 1$ and $\ell 2$), see Figure 5d. The choice of length scale therefore affects how far away two spatial field values need to be in $\boldsymbol{r}$-space before they can be significantly different. Furthermore, the spatial fields $g_1(\boldsymbol{r},\omega)$ and $g_2(\boldsymbol{r},\omega)$, representing the normalized strain to failure and strength, are assumed spatially correlated. Therefore, the cross-covariance function $\mathrm{cov}(g_1(\boldsymbol{r},\omega), g_2(\boldsymbol{r}',\omega))$ needs to be modeled to describe this spatial correlation. The cross-covariance is modeled as

$$K_{12}(\boldsymbol{r},\boldsymbol{r}') = \rho_{12}\exp\left(-\frac{|\boldsymbol{r}-\boldsymbol{r}'|^2}{2\ell_{12}{}^2}\right), \tag{15}$$

i.e., using an exponential kernel function with a correlation coefficient $\rho_{12}$ and characteristic length $\ell_{12}$. A total of four unknowns $\ell_1$, $\ell_2$, $\rho_{12}$ and $\ell_{12}$, need to be determined based on the one dimensional sample spatial field data in Figure 5d. For a discretized random field, the multivariate covariance matrix is given by

$$\boldsymbol{K} = \begin{bmatrix} \boldsymbol{K}_{11} & \boldsymbol{K}_{12} \\ \boldsymbol{K}_{12} & \boldsymbol{K}_{22} \end{bmatrix}, \tag{16}$$

where $\boldsymbol{K}_{11,pq} = K_{11}(\boldsymbol{r}_p,\boldsymbol{r}_q)$, $\boldsymbol{K}_{22,pq} = K_{22}(\boldsymbol{r}_p,\boldsymbol{r}_q)$, and $\boldsymbol{K}_{12,pq} = K_{12}(\boldsymbol{r}_p,\boldsymbol{r}_q)$. Using the one-dimensional spatial field data, see Figure 5d, these matrices become $N \times N$ covariance matrices. The sample data from both random fields can be collected into the vector

$$\boldsymbol{g}(\omega) = \begin{bmatrix} \boldsymbol{g}_1(\omega)^T & \boldsymbol{g}_2(\omega)^T \end{bmatrix}^T \tag{17}$$

where $\boldsymbol{g}_1(\omega) = \begin{bmatrix} g_1(x_1,\omega) & \cdots & g_1(x_N,\omega) \end{bmatrix}^T$ and $\boldsymbol{g}_2(\omega) = \begin{bmatrix} g_2(x_1,\omega) & \cdots & g_2(x_N,\omega) \end{bmatrix}^T$. This set of observations $\boldsymbol{g}(\omega)$ can be regarded as samples from a multivariate Gaussian distribution, i.e., $\boldsymbol{g}(\omega) \sim \mathcal{N}(\boldsymbol{0},\boldsymbol{K})$. The log-marginal likelihood function [38] (MLII) is the natural logarithm of the distribution of $\boldsymbol{g}(\omega)$, which can be written as

$$\log\mathcal{N}(\boldsymbol{0},\boldsymbol{K}) = -\frac{1}{2}\left(\boldsymbol{g}(\omega)^T\boldsymbol{K}^{-1}\boldsymbol{g}(\omega) + \log|\boldsymbol{K}| + 2N\log 2\pi\right). \tag{18}$$

A widely used method for the determination of unknown parameters of the Gaussian random fields is the maximization of the log-marginal likelihood function. In this work the unknown parameters $\ell_1$, $\ell_2$, $\rho_{12}$, and $\ell_{12}$ are determined by the maximization of the log-marginal likelihood function according to Equation (18) subject to three constraints. This can equivalently be written as a minimization problem according to

$$\begin{cases} \min\limits_{\ell_1,\, \ell_2,\, \rho_{12},\, \ell_{12}} \quad \boldsymbol{g}(\omega)^T \boldsymbol{K}^{-1} \boldsymbol{g}(\omega) + \log|\boldsymbol{K}| \\ \text{s.t. } \rho_{\varepsilon_f \sigma_f} = \rho^t_{\varepsilon_f \sigma_f} \\ \text{s.t } \mu_{\text{upcr1}} = \mu^t_{\text{upcr1}} \\ \text{s.t } \mu_{\text{upcr2}} = \mu^t_{\text{upcr2}} \end{cases} \tag{19}$$

The first constraint enforces that correlation coefficient between the strength and strain to failure $\rho_{\varepsilon_f \sigma_f}$ matches the one computed from the analysis of the $L \times W_{\text{SVE}}$ specimen in Figure 5d, where $\rho^t_{\varepsilon_f \sigma_f}$ is given by Equation (9). The second and third constraints enforces that the average distance between two zero levels up-crossings for the normalized spatial fields $g_1(x, \omega)$ and $g_2(x, \omega)$, respectively, matches the target ones, $\mu^t_{\text{upcr1}}$ and $\mu^t_{\text{upcr2}}$, computed in Equation (13).

Successful solution of the optimization problem according to Equation (19) is simplified if good initial guesses for the unknown parameters are used. The following initial values are used in the first iteration:

$$\begin{cases} \ell_1^{(0)} = \frac{\mu^t_{\text{upcr1}}}{2\pi} \\ \ell_2^{(0)} = \frac{\mu^t_{\text{upcr2}}}{2\pi} \\ \rho_{12}^{(0)} = \rho^t_{\varepsilon_f \sigma_f} \\ \ell_{12}^{(0)} = \frac{\ell_1^{(0)} + \ell_2^{(0)}}{2}. \end{cases} \tag{20}$$

As is seen, the initial guesses for the characteristic length scales $\ell_1^{(0)}$ and $\ell_2^{(0)}$ are proportional to the target average distance between zero-level up-crossings. The proportionality constant $2\pi$ is based on the analysis of univariate one-dimensional spatial fields with an underlying exponential Gaussian covariance function [38]. Furthermore, the initial guess $\rho_{12}^{(0)}$ is given by $\rho^t_{\varepsilon_f \sigma_f}$. It should be observed that $\rho_{12} \neq \rho_{\varepsilon_f \sigma_f}$ due to the non-linear transformation that relates the normalized spatial fields and the original spatial fields. The initial guess $\ell_{12}^{(0)}$ is chosen arbitrarily.

In order to generate new realizations of 2D-spatial fields of strain to failure $\varepsilon_f(\boldsymbol{r}, \omega)$ and strength $\sigma_f(\boldsymbol{r}, \omega)$, simulation of Gaussian random fields $g_1(\boldsymbol{r}, \omega)$ and $g_2(\boldsymbol{r}, \omega)$ is first performed. The KL-expansion is used to simulate discrete and correlated Gaussian random fields $\boldsymbol{g}(\omega) = \begin{bmatrix} \boldsymbol{g}_1(\omega)^T & \boldsymbol{g}_2(\omega)^T \end{bmatrix}^T$, see Section 2.3. It can be written as

$$\boldsymbol{g}(\omega) = \sum_{\boldsymbol{i}=\boldsymbol{1}}^{\boldsymbol{N}} \sqrt{\lambda_i} \boldsymbol{\eta}_i \xi_i(\omega), \tag{21}$$

where $\{\lambda_i\}_{i=1}^N$ and $\{\boldsymbol{\eta}_i\}_{i=1}^N$ are eigenvalues and eigenvectors of the covariance matrix $\boldsymbol{K}$ in Equation (16). Since $\boldsymbol{g}(\omega) \sim \mathcal{N}(\boldsymbol{0}, \boldsymbol{K})$, the random variables $\xi_i(\omega)$ become independent standard normal variables. A transformation is thereafter performed according to

$$\begin{cases} \varepsilon_f(\boldsymbol{r}, \omega) = F_{\varepsilon_f}{}^{-1}\{\phi[g_1(\boldsymbol{r}, \omega)]\} \\ \sigma_f(\boldsymbol{r}, \omega) = F_{\sigma_f}{}^{-1}\{\phi[g_2(\boldsymbol{r}, \omega)]\} \end{cases} \tag{22}$$

### *3.3. Constitutive Model*

An empirical relation between the equivalent stress $\sigma_{\text{eq}}$ and the equivalent strain $\varepsilon_{\text{eq}}$ of the form

$$\frac{\sigma_{\text{eq}}}{\sigma_f} = \left(1 - f_1(\kappa, n)\left(\frac{\varepsilon_{\text{eq}}}{\varepsilon_f}\right)^n\right) f_2(\kappa, n) \tan\text{h}\left(\kappa \frac{\varepsilon_{\text{eq}}}{\varepsilon_f}\right), \quad \varepsilon_{\text{eq}} \leq \varepsilon_f \tag{23}$$

is assumed to model the response of each SVE in the network. In Equation (23), $\kappa > 0$ is a parameter controlling the elastic and elastoplastic modulus, $n > 0$ is a damage exponent and $f_{1-2} > 0$. The relation is a variant of a Bammann–Chiesa–Johnson (BCJ) model previously used in stochastic constitutive modeling [39]. Enforcing that $\sigma_{\text{eq}}(\varepsilon_{\text{eq}} = \varepsilon_{\text{f}}) = \sigma_{\text{f}}$ and $\frac{\partial \sigma_{\text{eq}}}{\partial \varepsilon_{\text{eq}}}\Big|_{\varepsilon_{\text{eq}}=\varepsilon_{\text{f}}} = 0$, yields the following relations

$$\begin{cases} f_1(\kappa, n) = \frac{\kappa}{n \sinh(\kappa) \cosh(\kappa) + \kappa} \\ f_2(\lambda, n) = \frac{n \sinh(\kappa) \cosh(\kappa) + \kappa}{n \sinh^2(\kappa)} \end{cases}. \tag{24}$$

The elastic modulus can be written as

$$E = \frac{n \sinh(\kappa) \cosh(\kappa) + \kappa}{n \sinh^2(\kappa)} \frac{\sigma_{\text{f}}}{\varepsilon_{\text{f}}}. \tag{25}$$

Since $f_{1-2}$ are functions of $n$ and $\kappa$, it is sufficient to study the effect of these two parameters in Equation (23). As can be seen from Figure 6, increasing the damage exponent $n$ or decreasing the parameter $\kappa$ has a similar effect on the stress–strain curve. In the proposed model, one value of the damage exponent $n$ is assumed suitable to model the response of SVEs of same size, while $\kappa$ is fitted to each SVE realization. The residual sum of squares resulting from fitting Equation (23) to a uniaxial test performed on the $j^{\text{th}}$ SVE is denoted by $e_j(n, \kappa_j)$. The optimal parameter $n$ is found by minimizing the total sum of these errors for all SVEs, i.e.,

$$\begin{cases} \min\limits_{n} \sum_{j=1}^{N_{\text{SVE}}} e_j\left(n, \kappa_j^*\right) \\ \text{s.t.} \quad \kappa_j^* = \arg \min\limits_{\kappa_j} e_j(n, \kappa_j), \quad j = 1..N_{\text{SVE}} \end{cases}, \tag{26}$$

where $N_{\text{SVE}}$ is the total number of SVEs and the parameters $\left\{\kappa_j^*\right\}_{j=1}^{N_{\text{SVE}}}$ are found by minimizing the error for each uniaxial test, as is seen from the constraints in Equation (26). The procedure is outlined in Figure 7. A large number of SVEs of the same size are randomly cut from a large fiber network, see Figure 7a. This assures that the SVEs are uncorrelated. A numerical uniaxial tensile test is performed on all extracted SVEs, see Figure 7b. All curves are scaled by their respective strength and strain to failure, see Figure 7c, and the relation in Equations (23) and (24) is fitted to the responses according to Equation (26).

From Figure 6b, it is apparent that increasing the parameter $\kappa$, for a constant $n$ (i.e., constant SVE size), also corresponds to an increasing elastic and plastic work in a uniaxial test. It is therefore clear that larger $\kappa$ values are typical for responses characterized by larger strain to failure values. This leads to the assumption $\kappa \propto \varepsilon_{\text{f}}$, see Figure 7d and Section 4.1. A stochastic relation of the form

$$\kappa = c_1 + c_2 \varepsilon_{\text{f}} + R \tag{27}$$

is assumed, where $c_1$ and $c_2$ are found by least squares fit for SVEs of the same size and the residual error $R \sim \mathcal{N}(0, s_R)$ is assumed to follow a normal distribution with a standard deviation $s_R$ computed from the residual fitting errors $\{R_j\}_{j=1}^{N_{\text{SVE}}}$, see Figure 7d. The last term in Equation (27) ensures that the elastic modulus according to Equation (25) is not solely dependent on $\sigma_{\text{f}}$ and $\varepsilon_{\text{f}}$ but involves an additional random component. The uncertainty can be seen as an epistemic model error, resulting from the assumption according to Equation (27).

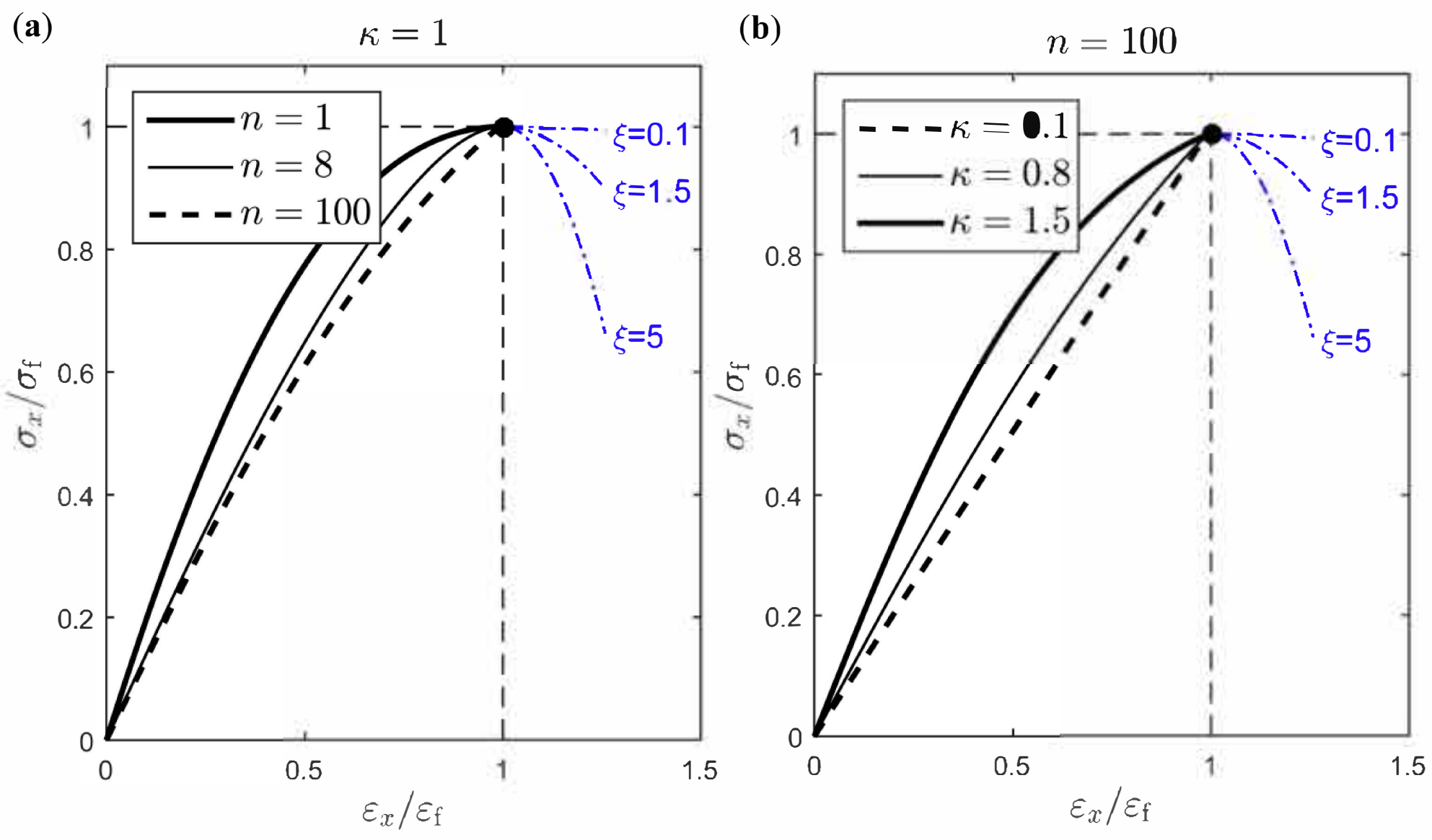

**Figure 6.** Effect on the stress–strain relation of (**a**) damage exponent $n$ and (**b**) parameter $\kappa$ in Equation (23).

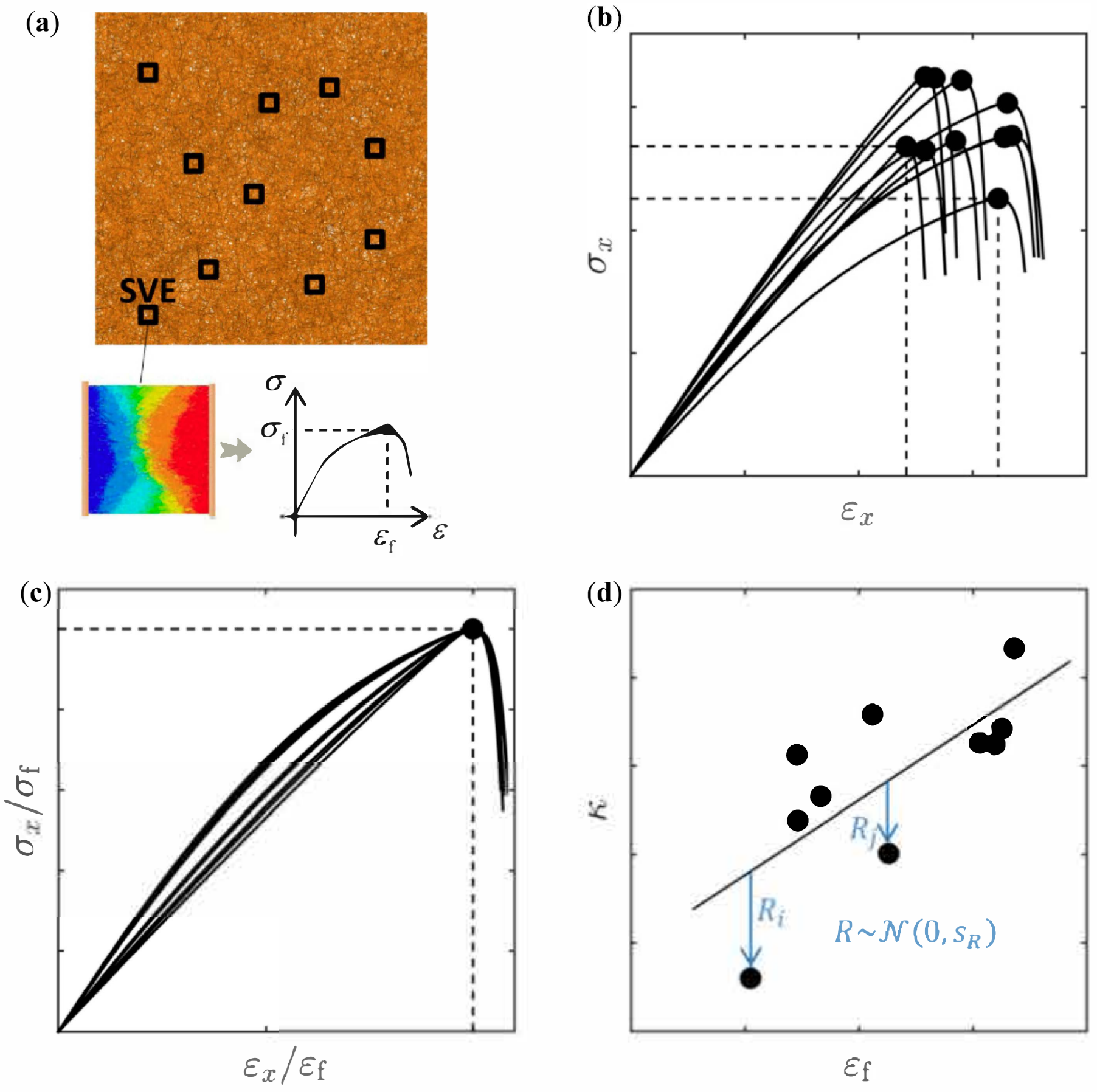

**Figure 7.** (**a**) SVEs of the same size randomly cut from a thin fiber network. (**b**) Constitutive response from uniaxial tensile tests on the SVEs. (**c**) Scaled constitutive response. (**d**) Parameter $\kappa$ as a function of $\varepsilon_f$ and residual error $R$.

The softening part of the SVE response is modeled as

$$\frac{\sigma_{\text{eq}}}{\sigma_{\text{f}}} = -\xi\left(\frac{\varepsilon_{\text{eq}}}{\varepsilon_{\text{f}}} - 1\right)^2 + 1, \quad \varepsilon_{\text{eq}} \geq \varepsilon_{\text{f}}, \tag{28}$$

where $\xi > 0$ is a parameter controlling the softening rate of the SVE response, see Figure 6. A small softening parameter $\xi = 0.15$ is chosen in order to increase the compliance of the system and assure numerical stability, but large enough for strain localization to onset in a large sample composed of SVEs.

An isotropic plane stress state is assumed to prevail, and a non-linear hardening model is used to model the constitutive response of the thin fiber network. The relation between the increment in stress vector $\sigma = \begin{bmatrix} \sigma_x & \sigma_y & \tau_{xy} \end{bmatrix}^T$ and strain vector $\varepsilon = \begin{bmatrix} \varepsilon_x & \varepsilon_y & \gamma_{xy} \end{bmatrix}^T$ is given by

$$\Delta\boldsymbol{\sigma} = \boldsymbol{D}_{\text{tan}}\Delta\boldsymbol{\varepsilon}, \tag{29}$$

where $\boldsymbol{D}_{\text{tan}} = \boldsymbol{D} - \boldsymbol{D}_{\text{pl}}$ is the tangent stiffness matrix. The elastic stiffness matrix is given by

$$\boldsymbol{D} = \frac{E}{1-\nu^2}\begin{bmatrix} 1 & \nu & 0 \\ \nu & 1 & 0 \\ 0 & 0 & \frac{1}{2}(1-\nu) \end{bmatrix}, \tag{30}$$

with Poisson's ratio $\nu = 0.3$ and elastic modulus given from Equation (25). The elastic–plastic tangent stiffness matrix is expressed as [40]

$$\boldsymbol{D}_{\text{pl}} = \frac{9E^2}{4A\sigma_{\text{eq}}{}^2(1-\nu^2)^2}\begin{bmatrix} s_{xx}^*{}^2 & s_{xx}^* s_{yy}^* & (1-\nu)s_{xx}^* s_{xy} \\ s_{yy}^* s_{xx}^* & \left(s_{yy}^*\right)^2 & (1-\nu)s_{yy}^* s_{xy} \\ (1-\nu)s_{xy}s_{xx}^* & (1-\nu)s_{xy}s_{yy}^* & (1-\nu)^2 s_{xy}{}^2 \end{bmatrix}, \tag{31}$$

where $\sigma_{\text{eq}}$ is the equivalent stress according to Equation (23) computed at the effective plastic strain $\varepsilon_{\text{e,pl}} = \left(\frac{2}{3}\sum_i\sum_j \varepsilon_{ij,\text{pl}}\varepsilon_{ij,\text{pl}}\right)^{1/2}$, $\varepsilon_{ij,\text{pl}}$ is the plastic strain component, $s_{ij}$ is the deviatoric stress component, $s_{xx}^* = s_{xx} + \nu s_{yy}$, $s_{yy}^* = s_{yy} + \nu s_{xx}$, and $A$ is given by

$$A = H + \frac{9E}{4(1+\nu)\sigma_{\text{eq}}{}^2}\left[\sum_i\sum_j s_{ij}s_{ij} + \frac{\nu}{1-\nu}\left(\sum_i s_{ii}\right)^2\right]. \tag{32}$$

In Equation (32), $H$ is the plastic modulus computed as $H = \left.\frac{\partial\sigma_{\text{eq}}}{\partial\varepsilon_{\text{eq}}}\right|_{\varepsilon_{\text{eq}}=\varepsilon_{\text{e,pl}}}$ using Equation (23). The consistency condition with a von Mises plasticity assumption states that the plastic strain increment is given by

$$\Delta\varepsilon_{ij,\text{pl}} = \Delta\varepsilon_{\text{e,pl}}\frac{3s_{ij}}{2\sigma_{\text{eq}}}. \tag{33}$$

The plane stress condition implies that $s_{zz} = (\sigma_x + \sigma_y)/3$ and therefore $\Delta\varepsilon_{z,\text{pl}} \neq 0$.

### 3.4. Finite-Element Implementation

The determined parameters $n$, $c_1$, $c_2$, and $s_R$ and the spatial fields $\varepsilon_{\text{f}}(x,y)$ and $\sigma_{\text{f}}(x,y)$ are used to specify the material properties in a Finite-Element (FE) model. The randomly generated spatial fields are applied on the integration points and extrapolated to the nodes. This ensures a continuous material property spatial field.

### 3.5. Summary of the Proposed Stochastic Constitutive Model

A flowchart of the proposed method is presented in Figure 8. The parameters $\ell_1$, $\ell_2$, $\rho_{12}$, and $\ell_{12}$ are used to construct the random spatial fields of strength and strain to failure $\sigma_{\mathrm{f}}(\boldsymbol{r},\omega)$ and $\varepsilon_{\mathrm{f}}(\boldsymbol{r},\omega)$. Given these spatial fields as input to the isotropic constitutive model in Section 3.3, as well as the parameters $n$, $c_1$, $c_2$, and $s_R$, the random tangent stiffness matrix $\boldsymbol{D}_{\tan}(\boldsymbol{r},\omega) = \boldsymbol{D}(\boldsymbol{r},\omega) - \boldsymbol{D}_{\mathrm{pl}}(\boldsymbol{r},\omega)$ can be determined. The randomness of the elastic stiffness matrix $\boldsymbol{D}$ given in Equation (30) at any spatial location $\boldsymbol{r}$ follows from the randomness of the elastic modulus. The latter can be expressed using Equations (25) and (27) according to

$$E(\boldsymbol{r},\omega) = \frac{n\sinh(c_1 + c_2\varepsilon_{\mathrm{f}}(\boldsymbol{r},\omega) + R)\cosh(c_1 + c_2\varepsilon_{\mathrm{f}}(\boldsymbol{r},\omega) + R)}{n\sinh^2(c_1 + c_2\varepsilon_{\mathrm{f}}(\boldsymbol{r},\omega) + R)}\frac{\sigma_{\mathrm{f}}(\boldsymbol{r},\omega)}{\varepsilon_{\mathrm{f}}(\boldsymbol{r},\omega)}. \quad (34)$$

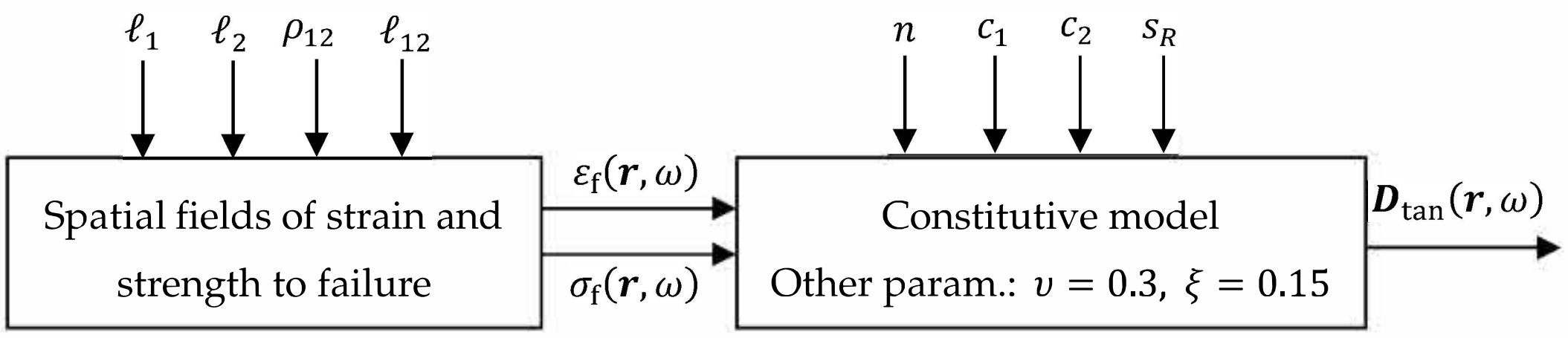

**Figure 8.** Flowchart of the proposed stochastic constitutive model.

As can be seen the parameter $s_R$ contributes to the random response through the normal random variable $R \sim \mathcal{N}(0, s_R)$. The same is true for the elastic-plastic tangent stiffness matrix. The parameters $\ell_1$, $\ell_2$, $\rho_{12}$, $\ell_{12}$ as well as $n$, $c_1$, $c_2$, and $s_R$ are all dependent on the chosen SVE size so that the constitutive response becomes independent of the latter. Poisson's ratio $\nu = 0.3$ is assumed constant and independent of SVE size. The softening behavior at each material point is assumed constant and negligible with $\xi = 0.15$.

## 4. Results

### 4.1. Determination of SVE Size-Dependent Stochastic Constitutive Model Parameters

The eight parameters in the stochastic constitutive model, $n$, $c_1$, $c_2$, $s_R$, $\ell_1$, $\ell_2$, $\rho_{12}$, and $\ell_{12}$ are dependent on the choice of SVE size, see Table 5. In this section, the determination of these parameters for an SVE size of $6 \times 6$ mm$^2$ and $4 \times 4$ mm$^2$ is demonstrated.

**Table 5.** Stochastic constitutive model parameters using $6 \times 6$ mm$^2$ or $4 \times 4$ mm$^2$ SVEs.

| SVE Size | $n$ | $c_1$ | $c_2$ | $s_R$ | $\ell_1$ | $\ell_2$ | $\rho_{12}$ | $\ell_{12}$ |
|---|---|---|---|---|---|---|---|---|
| $6 \times 6$ | 40.7 | 0.5058 | 0.7702 | 0.0671 | 0.85 | 1.95 | 0.55 | 1.25 |
| $4 \times 4$ | 34.6 | 0.4591 | 0.8017 | 0.0977 | 0.80 | 1.90 | 0.65 | 1.20 |

The parameters $n$, $c_1$, $c_2$, and $s_R$ in the relation according to Equation (23) and Equation (27) are first determined. A total of 70 SVEs of the same size are cut from random locations in a large fiber network, and a uniaxial test is performed on each SVE. The responses for all SVEs are shown in Figure 9a,b. The response of three SVEs together with the corresponding fitted model according to Equation (23) is shown in Figure 9c,d. Based on the procedure outline in Section 3.3, a value of $n = 40.7$ was found to minimize the sum of all least square errors from fitting Equation (23) to all $6 \times 6$ mm$^2$ numerical tests. The corresponding optimal $\kappa_j$ values, one for each SVE test, resulted in $c_1 = 0.5058$ and $c_2 = 0.7702$ after performing the linear regression $\kappa_j = c_1\varepsilon_{\mathrm{f}j} + c_2$, see Figure 9e. The standard deviation of the residual fitting error was $s_R = 0.0671$ and was used as a deterministic parameter in the stochastic relation according to Equation (27).

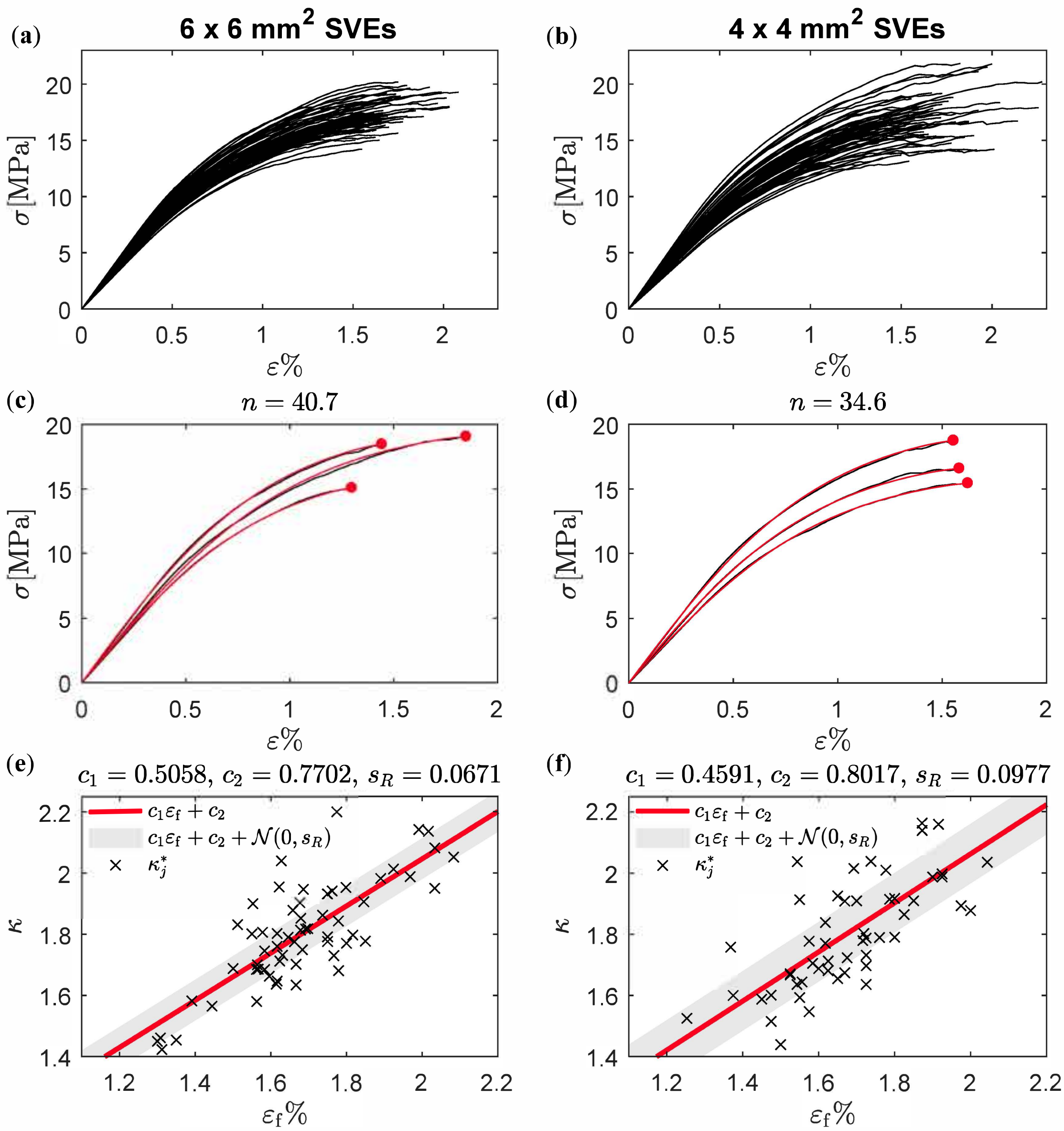

**Figure 9.** Seventy numerical experiments using direct simulations of uniaxially loaded (**a**) $6 \times 6$ mm$^2$ and (**b**) $4 \times 4$ mm$^2$ SVEs cut from different spatial locations in a fiber network. Response from direct simulation and fitted models for three (**c**) $6 \times 6$ mm$^2$ and (**d**) $4 \times 4$ mm$^2$ SVEs. Linear regression between optimal $\kappa$ values and $\varepsilon_f$ according to Equation (27) for (**e**) $6 \times 6$ mm$^2$ and (**f**) $4 \times 4$ mm$^2$ SVEs.

The random spatial field distributions of $\varepsilon_f(\boldsymbol{r}, \omega)$ and $\sigma_f(\boldsymbol{r}, \omega)$ were characterized by the parameters $\ell_1$, $\ell_2$, $\rho_{12}$, and $\ell_{12}$, see Table 5, which were determined by the optimization problem according Equation (19). In the following, the determination of the target values for the three constraints as well as initial guesses for the optimization parameters are detailed for the 6 mm $\times$ 6 mm SVE size, see Table 6.

**Table 6.** Constraints and the initial guess for the optimization problem according to Equation (19) for $6 \times 6$ mm$^2$ SVEs.

| Constraints | | | Initial Guess | | | |
|---|---|---|---|---|---|---|
| $\rho^t_{\varepsilon_f\sigma_f}$ | $\mu^t_{\text{upcr1}}$ | $\mu^t_{\text{upcr2}}$ | $\rho^{(0)}_{12}$ | $\ell^{(0)}_1$ | $\ell^{(0)}_2$ | $\ell^{(0)}_{12}$ |
| 0.55 | 6.3 | 11.9 | 0.55 | 1.0 | 1.9 | 1.45 |

Following the methodology outlined in Figure 5, a realization of the spatial fields $\varepsilon_f(x, \omega)$ and $\sigma_f(x, \omega)$ computed by discretizing a $360 \times 6$ mm$^2$ specimen using $6 \times 6$ mm$^2$ SVEs is shown in Figure 10a,b. A total of $N = 360$ points or SVE uniaxial tests are performed with a discretization distance of 1 mm. The target correlation $\rho^t_{\varepsilon_f\sigma_f}$ was determined by computing the correlation between $\varepsilon_f(x, \omega)$ and $\sigma_f(x, \omega)$, resulting in $\rho^t_{\varepsilon_f\sigma_f} = 0.55$. The initial guess for the parameter $\rho_{12}$ was given by $\rho^{(0)}_{12} = \rho^t_{\varepsilon_f\sigma_f} = 0.55$. In Figure 10c,d, the probability density functions of $\varepsilon_f$ and $\sigma_f$, computed from the 70 SVE tests used in the determination of $n$, $c_1$, $c_2$ and $s_R$ are shown together with the fitted kernel distributions.

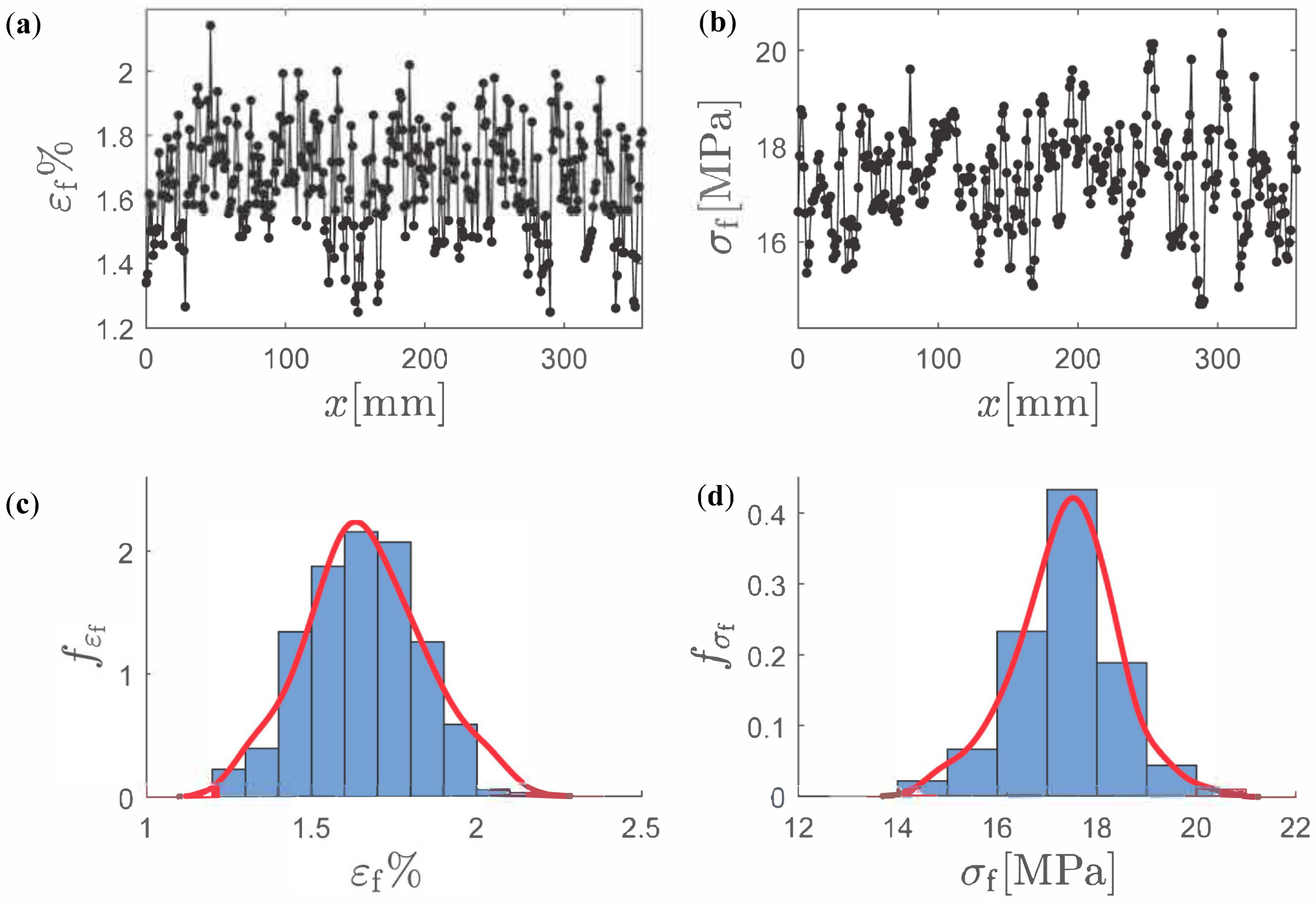

**Figure 10.** Spatial random fields of (**a**) $\varepsilon_f$ and (**b**) $\sigma_f$. Kernel approximation of the probability density function of (**c**) $\varepsilon_f$ and (**d**) $\sigma_f$.

The resulting normalized spatial fields $g_1(x, \omega) = \phi^{-1}\{F_{\varepsilon_f}[\varepsilon_f(x, \omega)]\}$ and $g_2(x, \omega) = \phi^{-1}\{F_{\sigma_f}[\sigma_f(x, \omega)]\}$, computed using the kernel approximation of the PDFs, are shown in Figure 11a,b. In Figure 11c,d, the average distance between two zero level up-crossings was computed for an increasing number of points along the length of the specimen. As can be seen, a specimen length of $L = 250$ mm was necessary for convergence to $\mu^t_{\text{upcr1}} = 6.3$ mm and $\mu^t_{\text{upcr2}} = 11.9$ mm. The initial guesses $\ell^{(0)}_1 = \mu^t_{\text{upcr1}}/2\pi = 1.0$ mm, $\ell^{(0)}_2 = \mu^t_{\text{upcr2}}/2\pi = 1.9$ mm, and $\ell^{(0)}_{12} = \frac{\ell^{(0)}_1 + \ell^{(0)}_2}{2} \approx 1.45$ can therefore be computed.

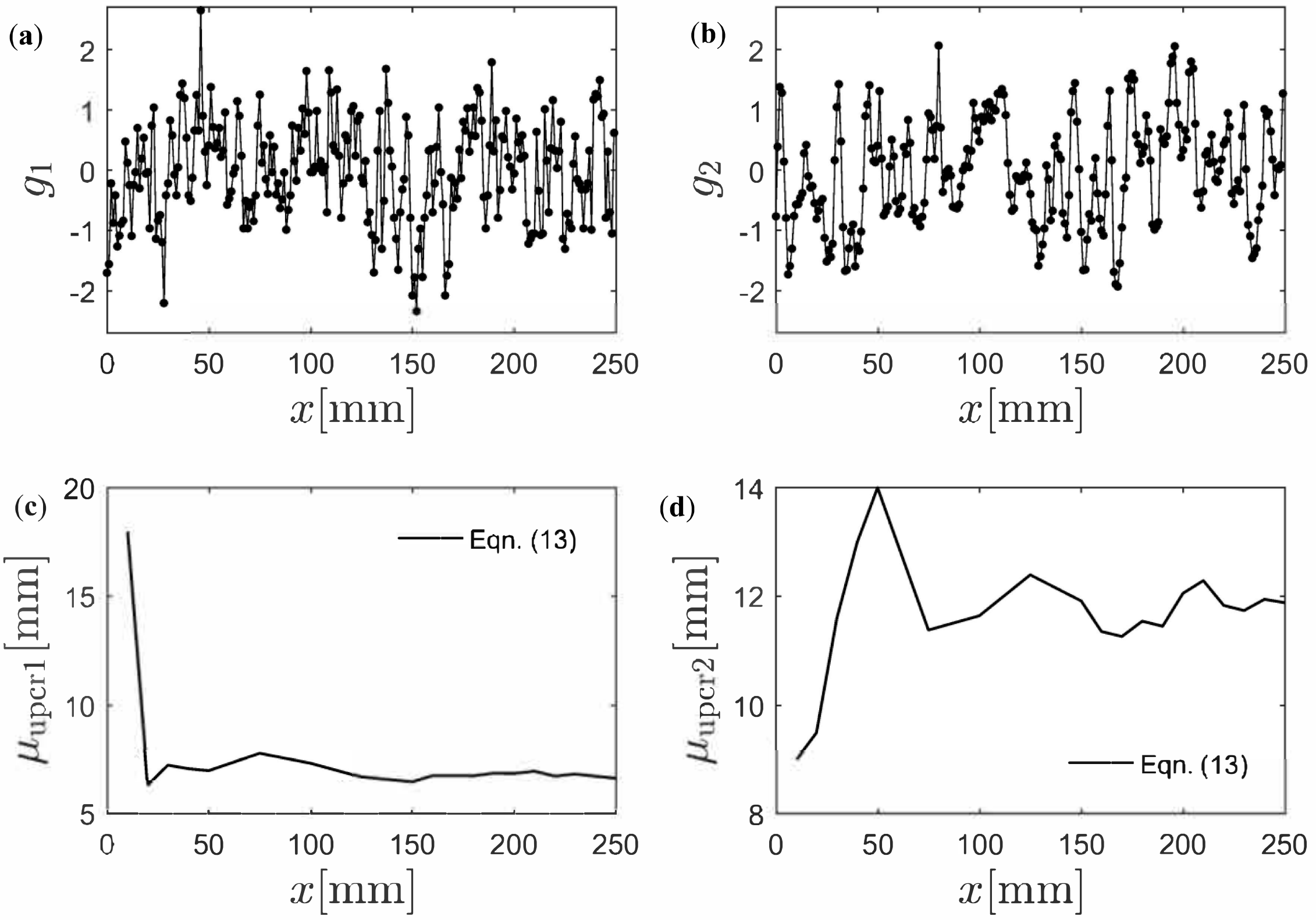

**Figure 11.** Transformed Gaussian random field of (**a**) $\varepsilon_{\mathrm{f}}$ and (**b**) $\sigma_{\mathrm{f}}$. Average distance between two zero-level up-crossings for (**c**) $g_1$ and (**d**) $g_2$ as a function of specimen length.

### *4.2. Continuum Reconstruction of A 18 × 18 mm$^2$ Sample and Choice of SVE Size*

Validation of the methodology was performed by comparing results from uniaxial tensile tests on the fiber network and the SVE-based continuum model. The boundary conditions applied on the fiber network is shown in Figure 12a. For the proposed continuum model, the same boundary conditions would result in stress concentration and in turn erroneous strain localization initiation. Therefore, the clamped ends in the fiber network were modeled in the continuum model using a contact boundary condition, see Figure 12b. Very low contact penalty stiffness was used to allow free contraction at the boundaries. In order to avoid uneven deformation at the ends due to the low contact stiffness, the boundary nodes were coupled with respect to their axial deformation $u_x$.

A uniaxial tensile test was performed on a specimen of size 18 × 18 mm$^2$ using direct simulation and the continuum model based on both a 4 × 4 mm$^2$ and a 6 × 6 mm$^2$ SVE size. The constitutive model parameters used are according to Table 5. The discretization distance (see Figure 5a), i.e., the distance between the centers of the SVEs, is half of the SVE size in both the length and width direction. The spatial distribution of stress $\sigma_{\mathrm{f}}$ and strain to failure $\varepsilon_{\mathrm{f}}$ for both SVE sizes is presented in Figure 13. As can be seen, the weakest material point using the 4 × 4 mm$^2$ SVE is at the lower right part of the specimen and has a value of 13.90 MPa. The corresponding value using 6 × 6 mm$^2$ SVEs is 14.86 MPa at approximately the same location. For the smaller SVE size, a higher variation in strength and strain to failure values can be observed within the specimen.

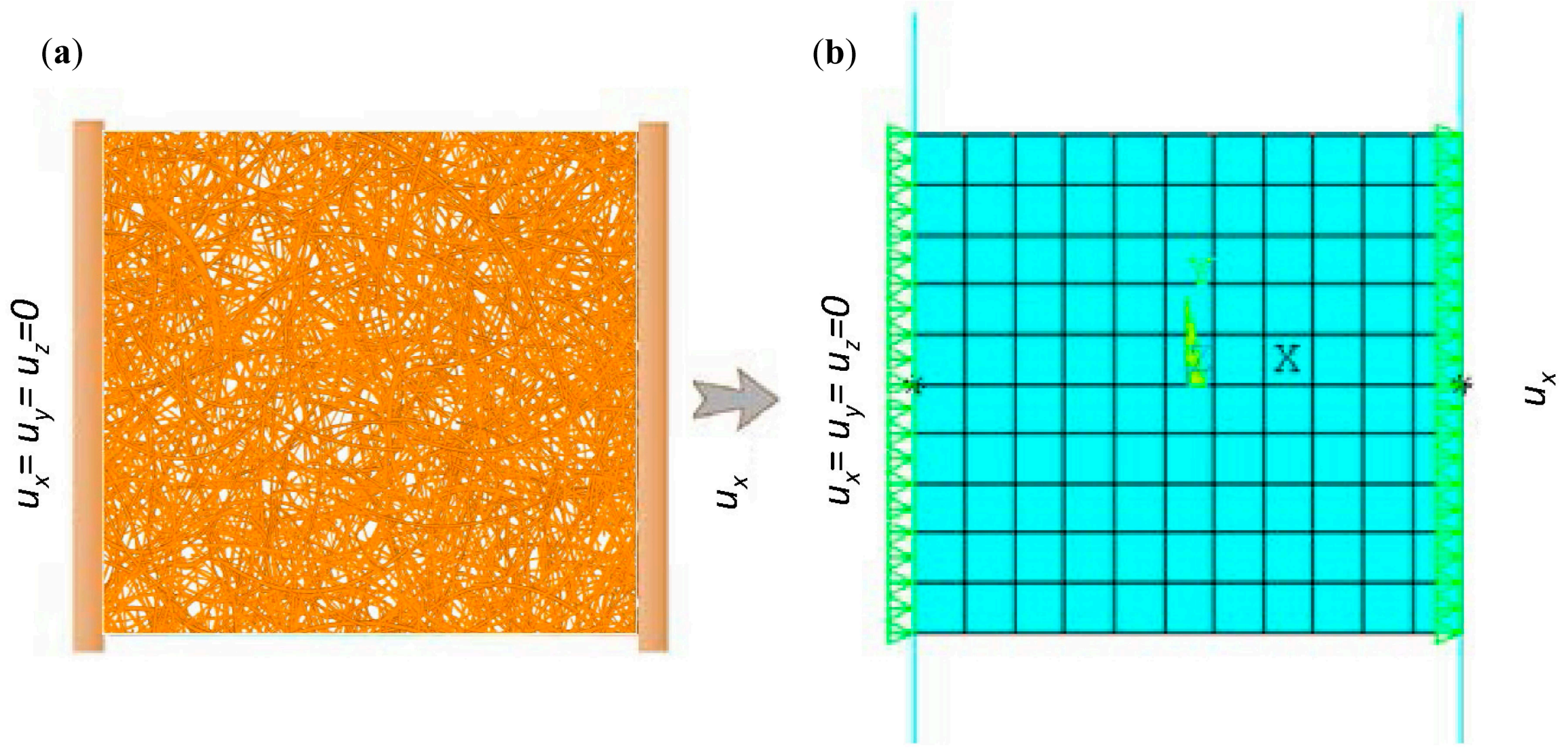

**Figure 12.** Finite element model and boundary conditions of a uniaxial test using (**a**) direct fiber simulation and (**b**) proposed continuum model.

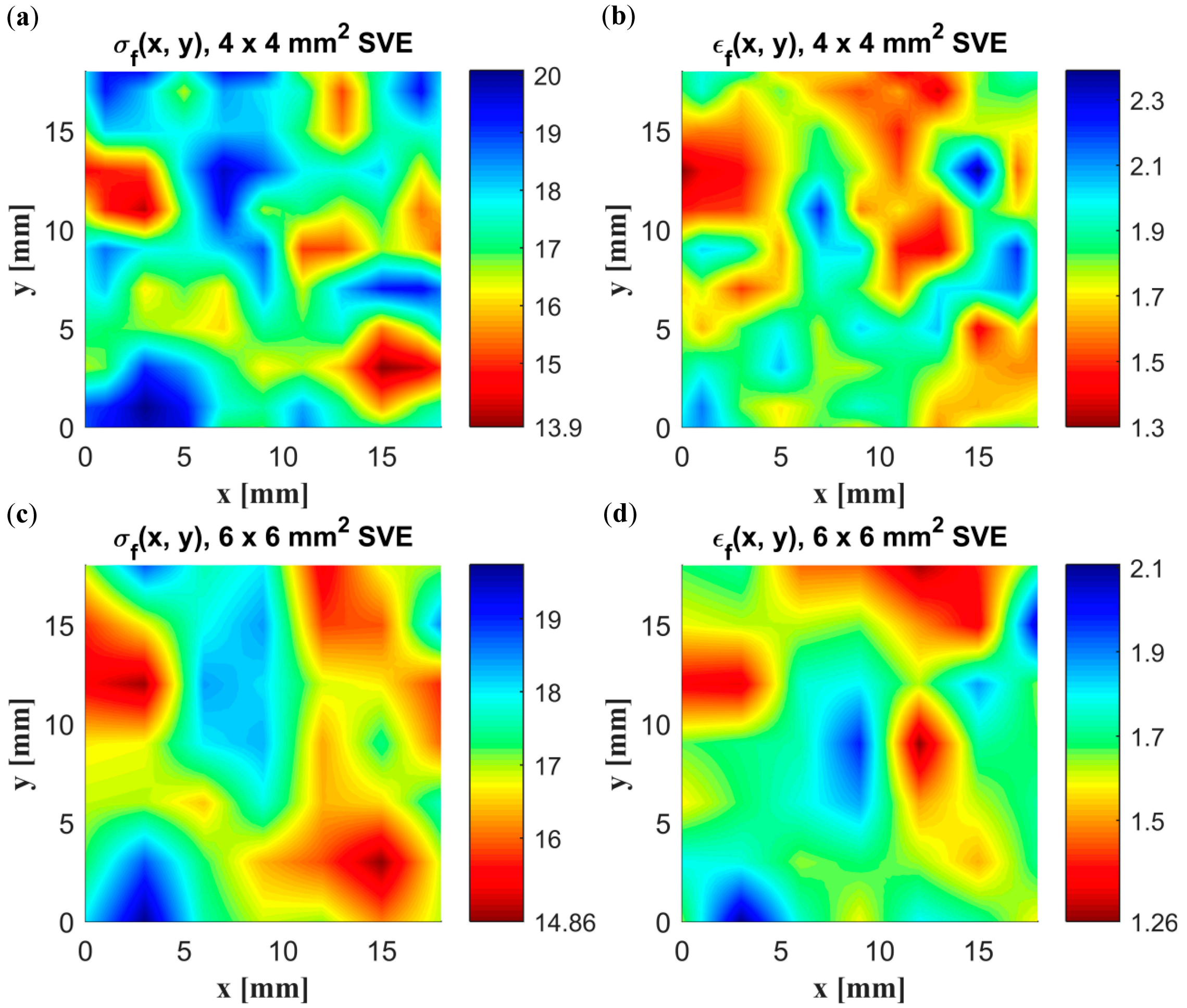

**Figure 13.** Spatial distribution of strength $\sigma_f$ (MPa) and strain to failure $\varepsilon_f$ (%) using SVEs of size (**a**) 4 × 4 mm$^2$ and (**b**) 6 × 6 mm$^2$ and a discretization distance of half the SVE size.

In Figure 14a, the uniaxial response in the $x$-direction using the continuum model is shown together with the response from the direct simulation. The results for both SVE sizes are in good agreement with direct simulation, with an absolute relative error in strength to failure of 0.4% and $-2\%$ for the $4 \times 4$ mm$^2$ and $6 \times 6$ mm$^2$ SVE size, respectively. The corresponding errors in strain to failure, 6% and $-6\%$, are, however, larger. In Figure 14b, the responses are also compared for an applied load in the transverse $y$-direction, with an overall performance similar to the $x$-direction.

In Figure 15, the spatial distribution of uniaxial strain in the loading direction is shown using direct simulation on the fiber network as well as the continuum model using $4 \times 4$ mm$^2$ and $6 \times 6$ mm$^2$ SVEs. The spatial distributions are shown for an increasing uniaxial load applied in the $x$-direction. By comparing Figures 13 and 14, it is clear that strain localization was initiated at weak material points. It is also seen that the strain localization pattern from direct simulation matched the ones from the continuum models relatively well.

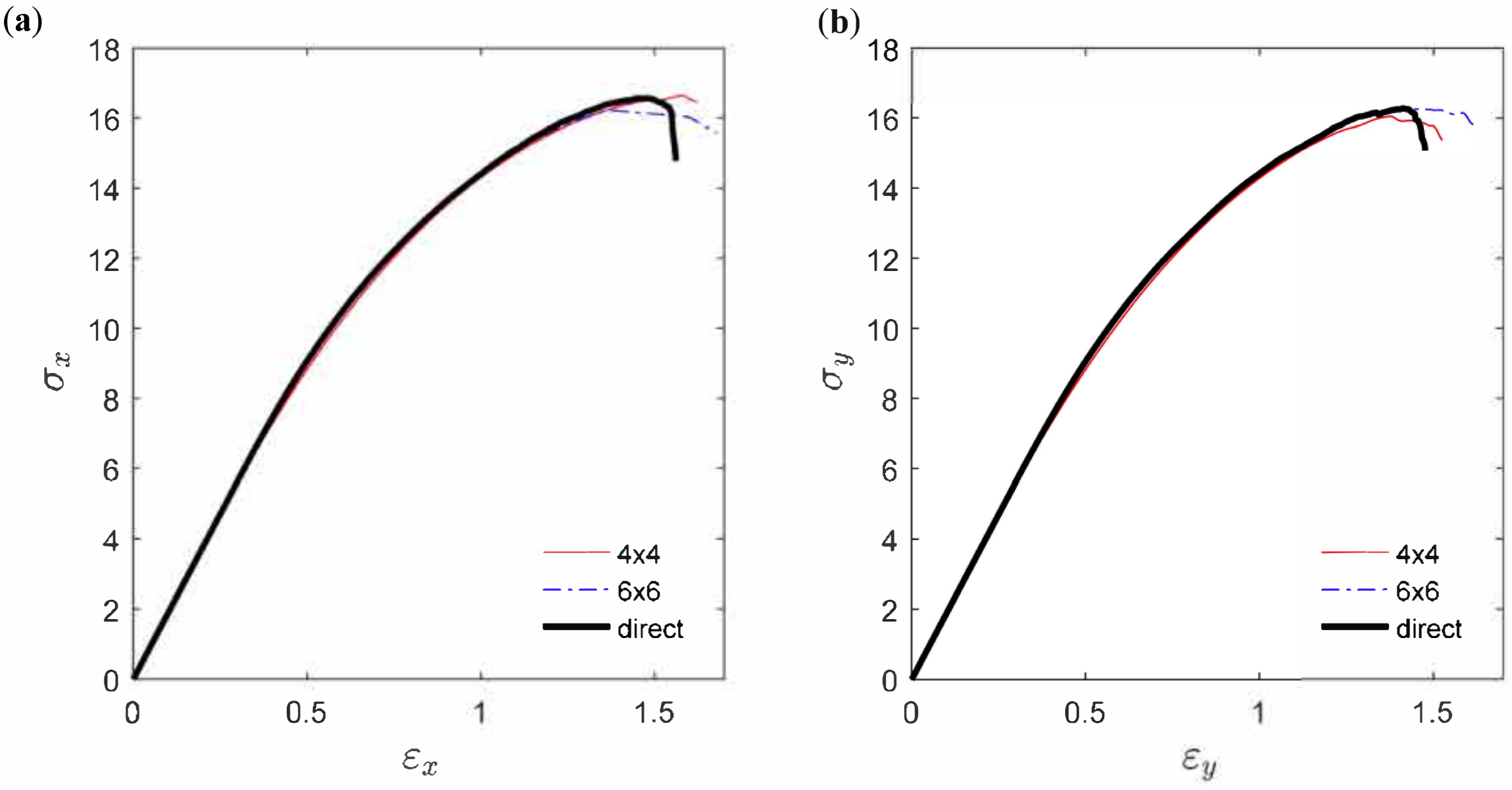

**Figure 14.** Uniaxial response in the (**a**) $x$-direction and (**b**) $y$-direction.

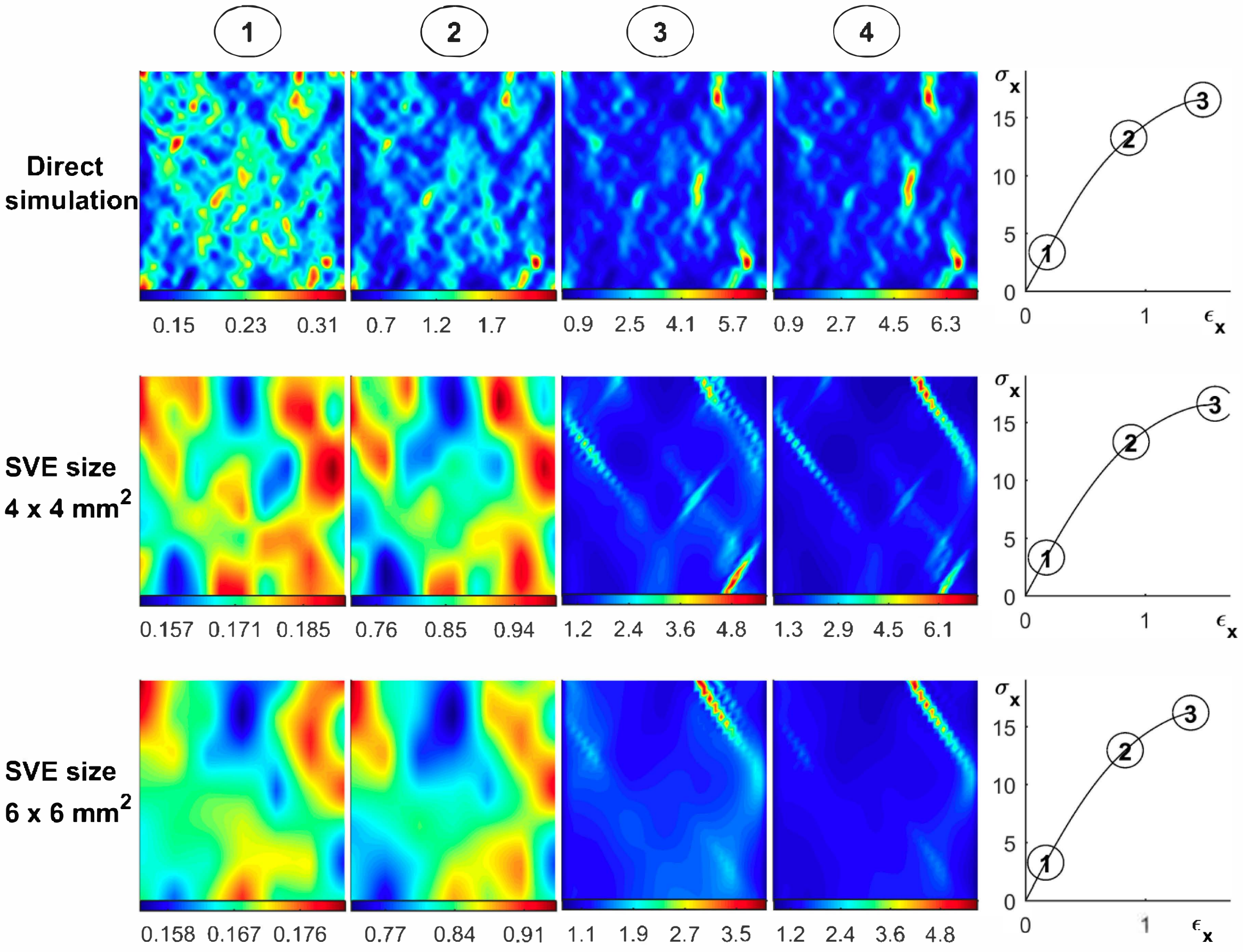

**Figure 15.** Spatial distribution of uniaxial strain with increased loading for: direct simulation, continuum model with SVE size 4 × 4 mm$^2$, and continuum model with SVE size 6 × 6 mm$^2$.

### *4.3. Random Continuum Simulation of 18 × 18 mm$^2$ and A4 Paper Sample*

In this section, random continuum realizations of the fiber network were subjected to a uniaxial loading. The proposed stochastic continuum model presented in Section 3 is used for both stochastic realization (Section 3.2) and constitutive response (Section 3.3). For a specimen size of 18 mm × 18 mm, the results are presented in Figure 16. Two correlated random spatial fields of strength and strain to failure, see Figure 16a,b, are first constructed based on an SVE size of 6 mm × 6 mm, i.e., using $\ell_1$, $\ell_2$, $\ell_{12}$, and $\rho_{12}$ according to Table 5. As can be seen, these randomly generated spatial fields are similar to the spatial fields obtained by cutting and testing direct fiber networks, see Figure 13c,d. The mechanical response is thereafter computed based on these spatial fields as well as the parameters $n$, $c_1$, $c_2$, and $s_R$, see Figure 16c. It should be emphasized that the randomness in the model is not only determined by $\sigma_f$ and $\varepsilon_f$ but also depends on $s_R$, as can be seen from the expression of the stiffness according to Equation (34). In Figure 16d, the uniaxial strain defined as the spatial derivative of the displacement field, is computed for the maximum applied stress marked in red in Figure 16c. As can be seen, damage localization was observed consistent with the observation from the analysis of the direct fiber network in Figure 15.

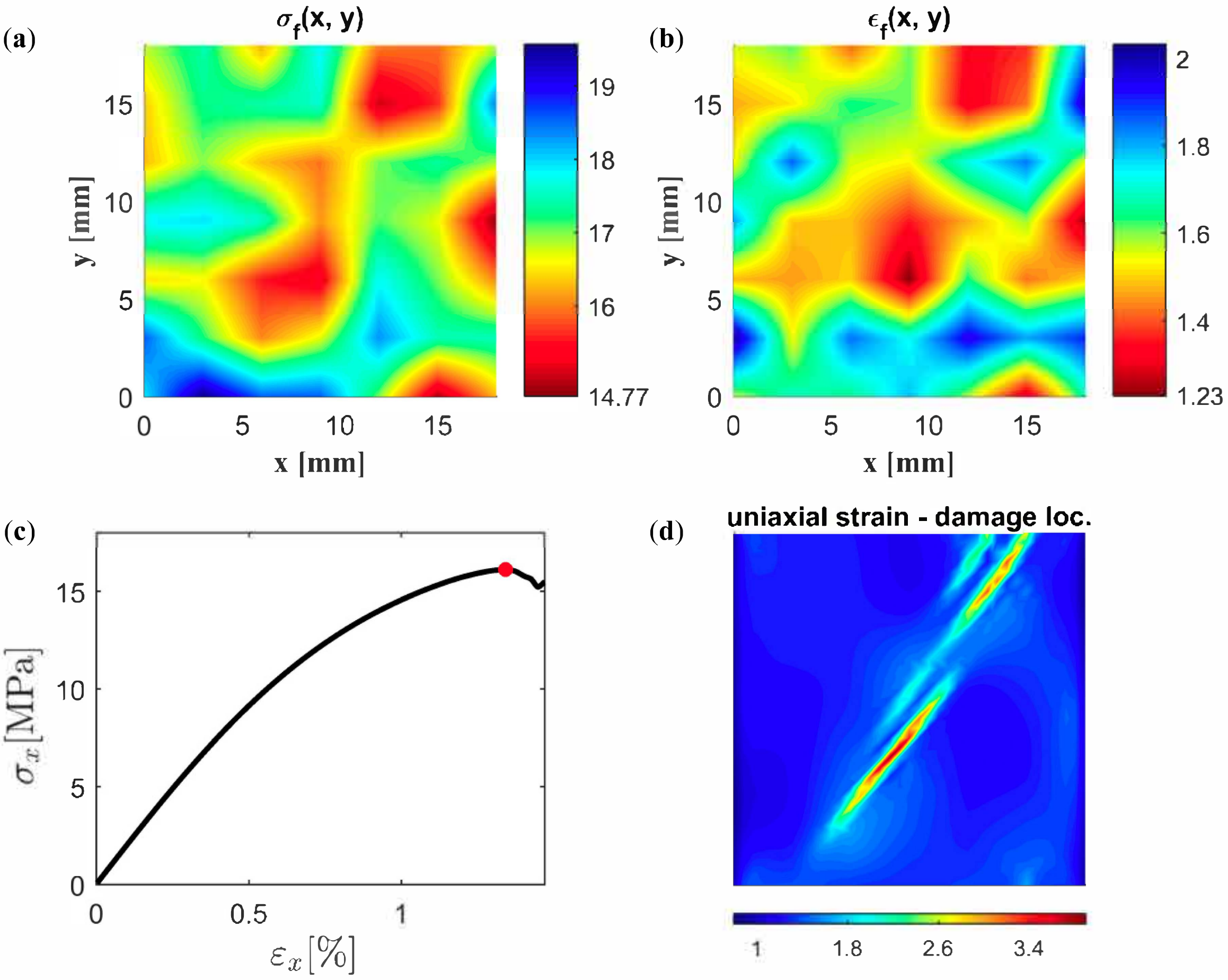

**Figure 16.** Randomly simulated spatial field of (**a**) strength $\sigma_f$(Mpa) and (**b**) strain to failure $\varepsilon_f$ (%); (**c**) uniaxial response; and (**d**) spatial distribution of uniaxial strain at the maximum applied stress (marked by a red point in (**c**)).

The major potential of the stochastic continuum model is demonstrated by the construction of larger specimens for which results cannot be validated using direct numerical simulation. A 297 × 210 mm$^2$ specimen (A4 paper size) was constructed and analyzed using the proposed model based on a 6 mm × 6 mm SVE size. Similarly to the result presented in Figure 16, the constructed spatial fields of strength and strain to failure are presented in Figure 17a,b, respectively, and the global uniaxial response of the specimen is presented in Figure 17c. The spatial distribution of uniaxial strain computed at the maximum global uniaxial stress, marked by a red point in Figure 17c, is presented in Figure 17d. By comparing Figures 17c and 16c, it is seen that the response is more brittle for the larger specimen. This is, for instance, seen by comparing the softening branch of the response in both figures.

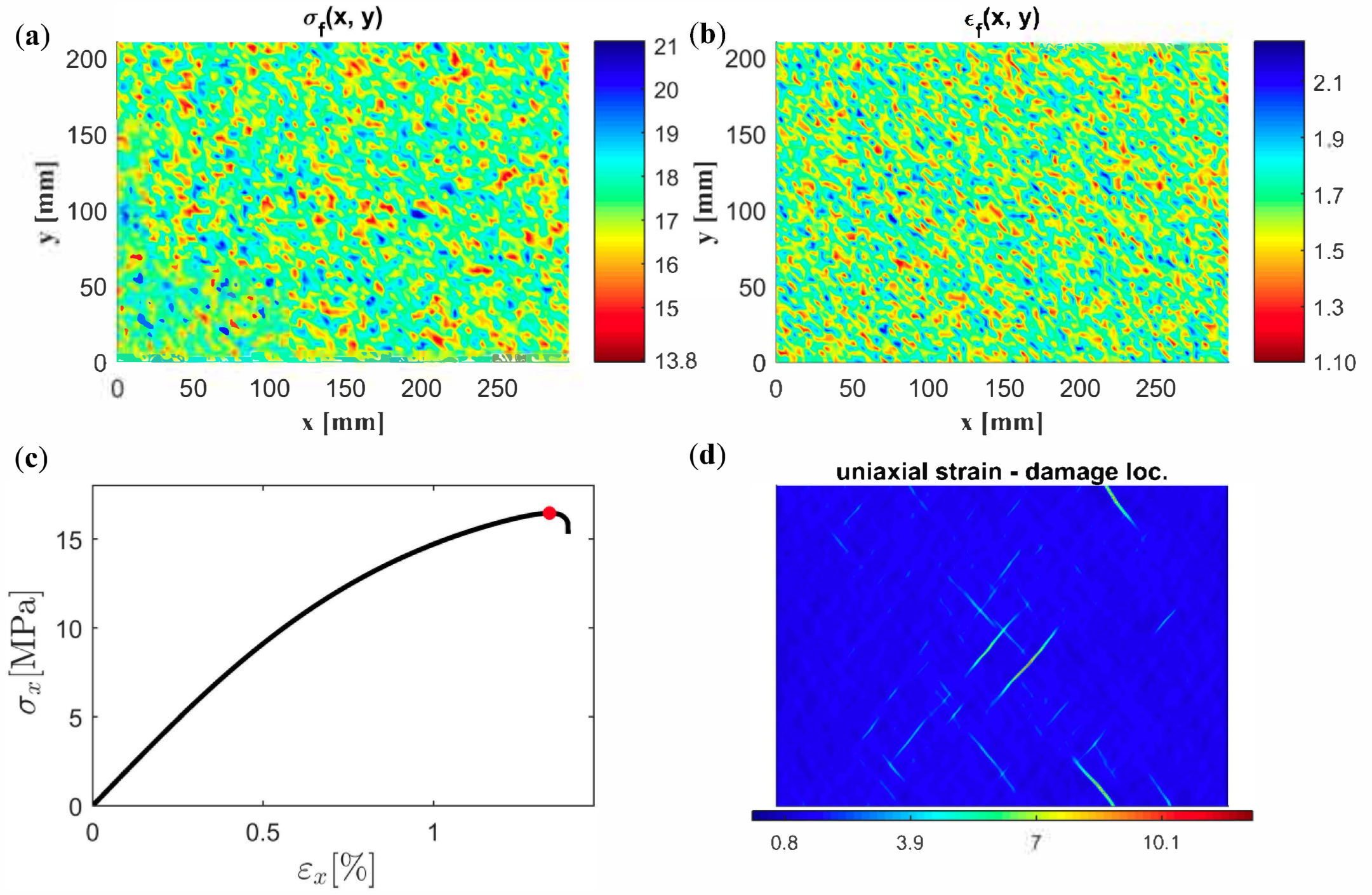

**Figure 17.** Randomly simulated spatial field of (**a**) strength $\sigma_f$ (Mpa) and (**b**) strain to failure $\varepsilon_f$ (%); (**c**) uniaxial response; and (**d**) spatial distribution of uniaxial strain at the maximum applied stress (marked by a red point in (**c**)).

### *4.4. Simulation of Stochastic Failure in Paper-Making Machines*

Consider a paper-making machine, Figure 18a, with a construction that results in breaks that mostly occur in an open draw of 1 m [18], see Figure 18b.

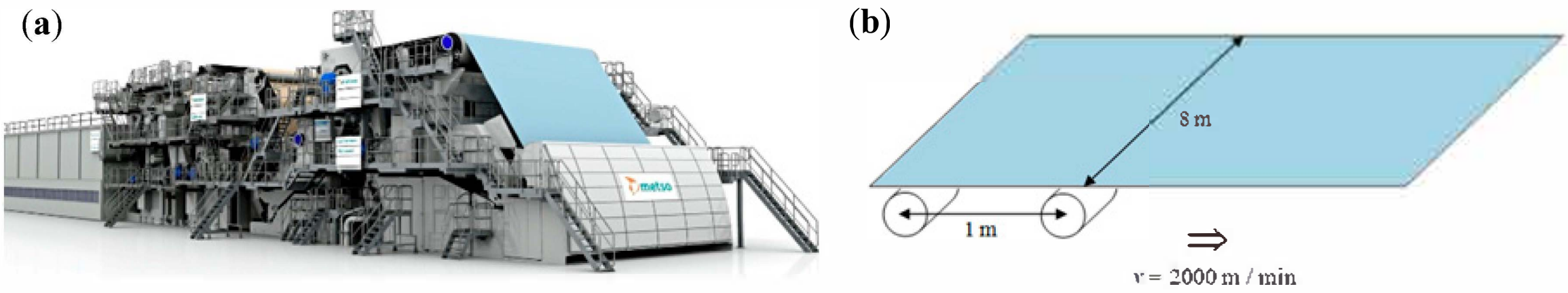

**Figure 18.** (**a**) Paper machine during the paper-making process. (**b**) Paperboard with rollers placed 1-m apart.

Using the proposed method, random realizations of 1 m × 8 m samples can be simulated. In Figure 19, three constructed spatial fields of strength and strain to failure are shown. It is seen that, for all three realizations, the difference between the strongest and weakest material points within the specimen is high due to the large considered size. From the uniaxial response also shown in the figure, a small variation in strength can be observed. The variation in corresponding strain to failure is, however, larger. This statistical variation in material response demonstrated using the proposed model, together with the variation in the applied load, can be seen as a major reason for the random occurrence of breaks in paper machines. From this example, it can be concluded that the disordered nature of the fiber network contributes to the statistical variation in mechanical response, even for such large specimens.

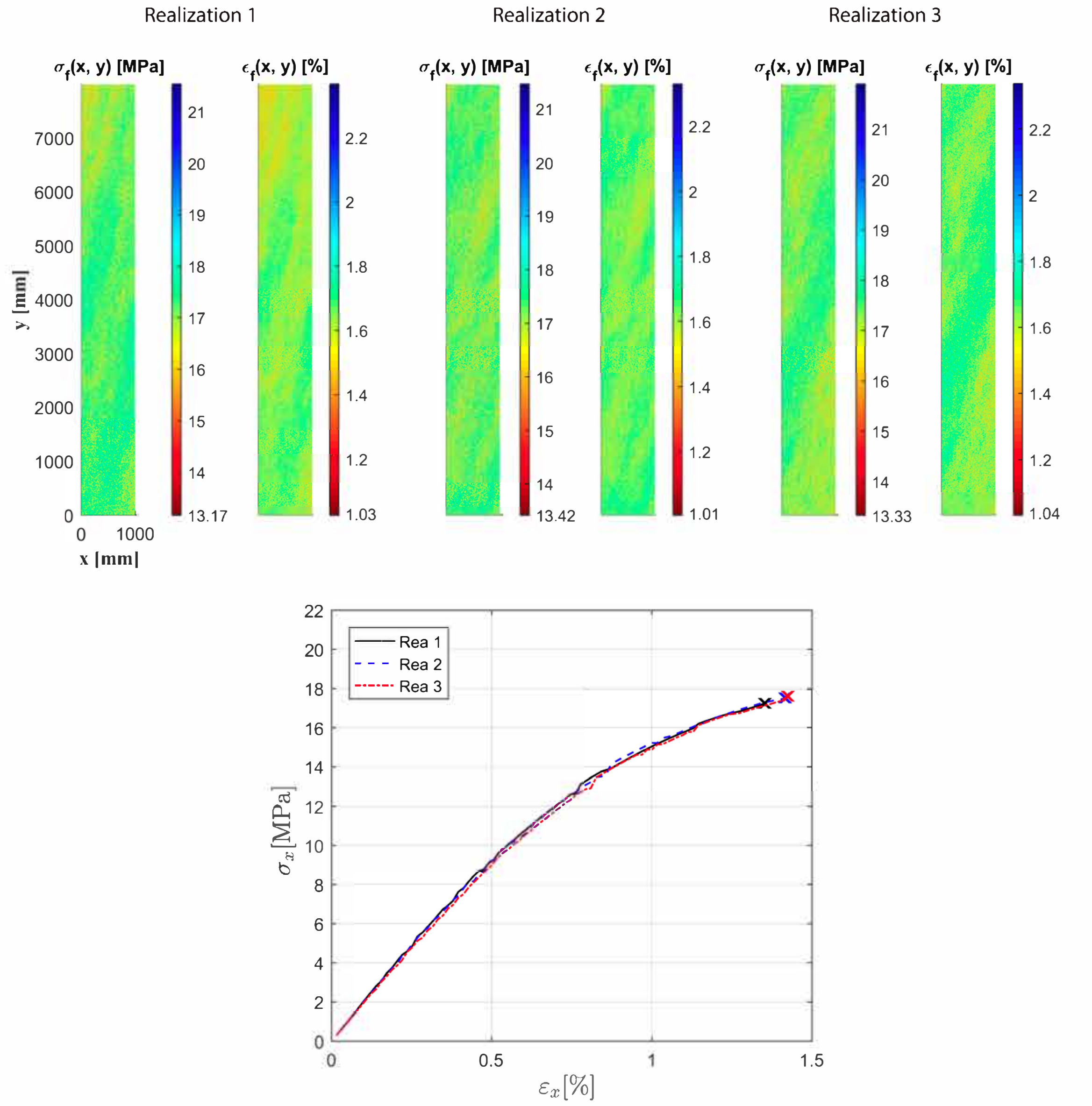

**Figure 19.** Three random realizations of 1 m × 8 m samples and corresponding uniaxial response.

## 5. Discussion

In this paper, a stochastic constitutive model for disordered fiber networks, derived through the mechanical response of stochastic volume element, is presented. The model is based on two correlated random spatial fields representing the spatial distribution of strength and strain to failure of the SVEs. These spatial fields are constructed using a multivariate kernel characterized by four parameters that are determined by analyzing realizations of direct fiber networks. The multiaxial non-linear mechanical response is modeled using an isotropic plasticity model with four additional parameters. The randomness in the material response is determined through the stochastic spatial fields $\varepsilon_{\mathrm{f}}(x,y)$ and $\sigma_{\mathrm{f}}(x,y)$ as well as the parameter $s_R$ in the constitutive model. All model parameters are dependent on the chosen SVE size. This assures that the global response is relatively independent of SVE size. The independence of SVE size is achieved as long as the ratio between the spatial correlation length and mesh size is large [22]. However, a large enough SVE should be chosen in order to limit the influence of boundary conditions during the numerical tests performed on SVEs cut from the direct fiber network. It should be noted that, the constitutive model obtained using SVEs face two sources of uncertainties: one contribution resulting from the applied boundary conditions and the other one from the uncertainties in the micro-structure. However as is seen in Reference [22], the uncertainties resulting from the micro-structure randomness, is assumed more important than the ones resulting from the applied boundary conditions.

The proposed methodology based on SVEs is validated by comparing the mechanical response of 18 mm × 18 mm samples using direct numerical simulation on the fiber network and the proposed continuum model. The results are compared through uniaxial loading in both the longitudinal and transverse direction. It has been shown that both mechanical response and strain localization pattern using the continuum model matches the one from the direct simulation relatively well. Based on the demonstrated accuracy, it may be concluded that using two spatial fields $\varepsilon_f(x, y)$ and $\sigma_f(x, y)$ is enough to characterize the material behavior and that the error from neglecting other spatial field variations is small. An important aspect explaining the error observed in the global strain and strength to failure is the assumption of isotropy of the SVE in the constitutive model. In future work, the use of an anisotropic plasticity model to describe the mechanical response of the SVE may further increase the accuracy in the global response. It should furthermore be noted that the results are mesh independent, since failure is assumed to occur when the maximum global stress is reached, and the mesh-dependent global softening behavior is therefore not of interest.

The proposed model paves the way to quantifying uncertainties in the response of thin fiber networks using the stochastic constitutive model. The eight parameters in the model can be linked to microstructural features [18] of the fiber network which opens the possibility to designing the network such that the variability in the global response is limited. For instance, it has been noted that the spatial fields of strain to failure and strength have different characteristic length. This is directly linked to microstructural features such as fiber connectivity and fiber orientation, which may affect the spatial fields differently. In the future, it is intended to enrich the present description by performing reliability estimation [41,42] and reliability-based design optimization (RBDO) [43], where both material and load variability are incorporated.

## 6. Conclusions

The conclusions in the papers can be summarized as follows.

- A stochastic constitutive model was proposed for isotropic fiber networks.
- The effect of fiber disorder for isotropic fiber networks can be quantified at the macromechanical level using the proposed method.
- In order to implement the stochastic constitutive model, eight SVE size-specific material parameters are necessary.
- The randomness in the model was simulated using random field realizations of strain to failure and strength. An epistemic uncertainty parameter also contributes to the randomness.
- Different characteristic lengths for the spatial fields of strain to failure and strength are assumed, which is in agreement with what is observed in direct fiber network simulation.
- The proposed model speeds up the simulation time of fiber networks compared to direct numerical simulations. A 24 mm × 24 mm simulation takes 2 min on a modern 128 GB RAM supercomputer while a corresponding direct fiber network simulation takes 2 days.
- The constitutive model captures the transition from ductile behavior for small sample size to semi-brittle behavior for larger sample size (Section 4.3).

**Supplementary Materials:** The software used for the fiber network reconstruction as well as the input data are available online at http://www.mdpi.com/1996-1944/12/3/538/s1.

**Author Contributions:** R.M.: Funding acquisition, Methodology, Software, Writing; A.K.: Funding acquisition, Methodology, Software, Supervision, Writing; W.C.: Supervision; M.O.: Methodology, Supervision.

**Funding:** The authors at the Royal Institute of Technology acknowledge and thank the VINN Excellence center BiMaC Innovation and its industrial partners as well as ÅForsk and The Swedish Research Council, grant number 2015-05282, for their financial support of this research. The computations resources were provided by the Swedish National Infrastructure for Computing (SNIC) at HPC2N, Umeå (Project SNIC2017-1-175). Wei Chen is supported by Center for Hierarchical Design (ChiMad NIST 70NANB14H012).

**Conflicts of Interest:** There is no conflict of interest.

## References

1. Uesaka, T. Statistical aspects of failure of paper products. In *Mechanics of Paper Products (Chapt 8)*; Niskanen, K., Ed.; De Gruyter: Berlin, Germany; Boston, MA, USA, 2012.
2. Hristopulos, D.T.; Uesaka, T. Structural disorder effects on the tensile strength distribution of heterogeneous brittle materials with emphasis on fiber network. *Phys. Rev. B* **2004**, *70*, 064108. [CrossRef]
3. Uesaka, T.; Juntunen, J. Time-dependent, stochastic failure of paper and box. *Nord. Pulp Pap. Res. J.* **2012**, *27*, 370–374. [CrossRef]
4. Mattsson, A.; Uesaka, T. Characterisation of time-dependent, statistical failure of cellulose fibre network. *Cellulose* **2018**, *25*, 2817–2828. [CrossRef]
5. Kulachenko, A.; Uesaka, T. Direct simulations of fiber network deformation and failure. *Mech. Mater.* **2012**, *51*, 1–14. [CrossRef]
6. Borodulina, S.; Motamedian, H.R.; Kulachenko, A. Effect of fiber and bond strength variations on the tensile stiffness and strength of fiber networks. *Int. J. Solids Struct.* **2018**, *154*, 19–32. [CrossRef]
7. Borodulina, S.; Kulachenko, A.; Tjahjanto, D.D. Constitutive modeling of a paper fiber in cyclic loading applications. *Comput. Mate. Sci.* **2015**, *110*, 227–240. [CrossRef]
8. Borodulina, S.; Nygårds, M.; Kulachenko, A. Stress-strain curve of paper revisited. *Nord. Pulp Pap. Res. J.* **2012**, *27*, 318–328. [CrossRef]
9. Li, Y.; Chen, Z.; Su, L.; Chen, W.; Jin, X.; Xu, H. Stochastic Reconstruction and Microstructure Modeling of SMC Chopped Fiber Composites. *Compos. Struct.* **2018**, *200*, 153–164. [CrossRef]
10. Chen, Z.; Huang, T.; Shao, Y.; Li, Y.; Xu, H.; Avery, K.; Zeng, D.; Chen, W.; Su, X. Multiscale finite element modeling of sheet molding compound (SMC) composite structure based on stochastic mesostructured reconstruction. *Compos. Struct.* **2018**, *188*, 25–38. [CrossRef]
11. Domaschke, S.; Zundel, M.; Mazza, E.; Ehret, A.E. A 3D computational model of electrospun networks and its application to inform a reduced modelling approach. *Int. J. Solid Struct.* **2018**, *158*, 76–89. [CrossRef]
12. Ban, E.; Barocas, V.H.; Shephard, M.S.; Picu, C.R. Effect of fiber crimp on the elasticity of random fiber network with and without embedding matrices. *J. Mech. Des.* **2016**, *83*, 041008. [CrossRef] [PubMed]
13. Godinho, P.M.J.S.; Jajcinovic, M.; Wagner, L.; Vass, V.; Fischer, W.J.; Bader, T.K.; Ulrich, H.; Bauer, W.; Eberhardsteiner, J.; Hellmich, C. A continuum micromechanics approach to the elasticity and strength of planar fiber networks: Theory and application to paper sheets. *Eur. J. Mech. A Solids* **2018**. In Press. [CrossRef]
14. Beex, L.A.A.; Peerlings, R.H.J.; Geers, M.G.D. A multiscale quasicontinuum method for lattice models with bond failure and fiber sliding. *Comput. Methods Appl. Mech. Eng.* **2014**, *269*, 108–122. [CrossRef]
15. Marulier, C.; Dumont, P.J.J.; Orgéas, L.; Caillerie, D.; Rolland du Roscoat, S. Towards 3D analysis of pulp fibre networks at the fibre and bond levels. *Nord. Pulp Pap. Res. J.* **2012**, *27*, 245–255. [CrossRef]
16. Bakar, I.A.A.; Kramer, O.; Bordas, S.; Rabczuk, T. Optimization of elastic properties and weaving patterns of woven composites. *Compos. Struct.* **2013**, *100*, 575–591. [CrossRef]
17. Wang, W.; Dai, Y.; Zhang, C.; Gao, X.; Zhao, M. Micromechanical Modeling of Fiber-Reinforced Composites with Statistically Equivalent Random Fiber Distribution. *Materials* **2016**, *9*, 624. [CrossRef] [PubMed]
18. Yin, X.; Chen, W.; To, A.; McVeigh, C.; Liu, W.K. Statistical volume element method for predicting microstructure-constitutive property relations. *Comput. Methods Appl. Mech. Eng.* **2008**, *197*, 3516–3529. [CrossRef]
19. Xu, H.; Greene, M.S.; Deng, H.; Dikin, D.; Brinson, C.; Liu, W.K.; Burkhart, C.; Papakonstantopoulos, G.; Poldneff, M.; Chen, W. Stochastic reassembly strategy for managing information complexity in heterogeneous materials analysis and design. *J. Mech. Des.* **2013**, *135*, 101010. [CrossRef]
20. Liu, Y.; Greene, M.S.; Chen, W.; Dikin, D.A.; Liu, W.K. Computational microstructure characterization and reconstruction for stochastic multiscale material design. *Comput.-Aided Des.* **2013**, *45*, 65–76. [CrossRef]

21. Hu, A.; Li, X.; Ajdari, A.; Jiang, B.; Burkhart, C.; Chen, W.; Brinson, C. Computational analysis of particle reinforced viscoelastic polymer nanocomposites—statistical study of representative volume element. *J. Mech. Phys. Solids* **2018**, *114*, 55–74. [CrossRef]
22. Lucas, V.; Golinval, J.-C.; Paquay, S.; Nguyen, V.-D.; Noels, L.; Wu, L. A stochastic computational multiscale approach; Application to MEMS resonators. *Comput. Methods Appl. Mech. Engrg.* **2015**, *294*, 141–167. [CrossRef]
23. Motamedian, H.R.; Kulachenko, A. Rotational Constraint between Beams in 3-D Space. *Mech. Sci.* **2018**, *9*, 373–387. [CrossRef]
24. Ibrahimbegovic, A. On finite element implementation of geometrically nonlinear Reissner's beam theory: Three-dimensional curved beam elements. *Comput. Methods Appl. Mech. Eng.* **1995**, *122*, 11–26. [CrossRef]
25. Stapleton, S.E.; Appel, L.; Simon, J.-W.; Reese, S. Representative volume element for parallel fiber bundles: Model and size convergence. *Compos. Part A Appl. Sci. Manuf.* **2016**, *87*, 170–185. [CrossRef]
26. Sonon, B.; Massart, T.J. A Level-Set Based Representative Volume Element Generator and XFEM Simulations for Textile and 3D-Reinforced Composites. *Materials* **2013**, *6*, 5568–5592. [CrossRef] [PubMed]
27. Geers, M.G.D.; Kouznetsova, V.G.; Matouš, K.; Yvonnet, J. Homogenization Methods and Multiscale Modeling: Nonlinear Problems. In *Encyclopedia Computational Mechanics*, 2nd ed.; Wiley Online Library: Hoboken, NJ, USA, 2017.
28. Sun, W.C.; Andrade, J.E.; Rudnicki, J.W. Multiscale method for characterization of porous microstructures and their impact on macroscopic effective permeability. *Int. J. Numer. Meth. Eng.* **2011**, *88*, 1260–1279. [CrossRef]
29. Gitman, I.M.; Askes, H.; Sluys, L.J. Representative Volume: Existence and size determination. *Eng. Fract. Mech.* **2007**, *74*, 2518–2534. [CrossRef]
30. Hashin, Z. Analysis of Composite Materials—A Survey. *J. Appl. Mech.* **1983**, *50*, 481–505. [CrossRef]
31. Ghanem, R.G.; Spanos, P.D. *Stochastic Finite Elements: A Spectral Approach*; Springer: New York, NY, USA, 1991.
32. Chen, W.; Yin, X.; Lee, S.; Liu, W.K. A multiscale design methodology for hierarchical systems with random uncertainties. *J. Mech. Des.* **2010**, *132*, 041006. [CrossRef]
33. Yin, X.; Lee, S.; Chen, W.; Liu, W.K. Efficient Random Field Uncertainty Propagation in Design Using Multiscale Analysis. *J. Mech. Des.* **2009**, *131*, 021006. [CrossRef]
34. Ghanem, R.G.; Doostan, A. On the construction and analysis of stochastic models: Characterization and propagation of the errors associated with limited data. *J. Comput. Phys.* **2006**, *217*, 63–81. [CrossRef]
35. Clément, A.; Soize, C.; Yvonnet, J. Uncertainty quantification in computational stochastic multiscale analysis of nonlinear elastic materials. *Comput. Methods Appl. Mech. Eng.* **2013**, *254*, 61–82. [CrossRef]
36. Mansour, R.; Olsson, M. Response surface single loop method with higher order reliability assessment. *Struct. Multidiscip. Optim.* **2016**, *54*, 63–79. [CrossRef]
37. Popescu, R.; Deodatis, G.; Prevost, J.H. Simulation of homogenous nonGaussian stochastic vector fields. *Probab. Eng. Mech.* **1998**, *13*, 1–13.
38. Rassmussen, C.E.; Williams, C.K.I. *Gaussian Process for Machine Learning*; The MIT Press: London, UK, 2006.
39. Greene, M.S.; Liu, Y.; Chen, W.; Liu, W.K. Computational uncertainty analysis in multiresolution materials via stochastic. *Comput. Methods Appl. Mech. Eng.* **2011**, *200*, 309–325. [CrossRef]
40. Ottosen, N.; Ristinmaa, M. *The Mechanics of Constitutive Modelling*; Elsevier: London, UK, 2005.
41. Mansour, R.; Olsson, M. Efficient reliability assessment with the conditional probability method. *J. Mech. Des.* **2018**, *140*, 081402. [CrossRef]
42. Mansour, R.; Olsson, M. A Closed-Form Second-Order Reliability Method Using Noncentral Chi-Squared Distributions. *J. Mech. Des.* **2014**, *136*, 101402. [CrossRef]
43. Aoues, Y. A Chateauneuf, Benchmark study of numerical methods for reliability-based design optimization. *Struct. Multidiscip. Optim.* **2010**, *41*, 277–294. [CrossRef]